\documentclass[reprint,aps,amsmath,amssymb,prfluids,onecolumn,11pt]{revtex4-2}
\usepackage{tabularx} 
\usepackage{graphicx} 
\usepackage{bm}       
\usepackage{natbib}   
\usepackage{tikz}     
\usepackage{amsmath}  
\usepackage{amssymb}  
\usepackage[normalem]{ulem}     
\usepackage{subcaption} 
\usepackage{epstopdf,epsfig}
\usepackage{textalpha}
\usepackage{setspace} 
\usepackage[normalem]{ulem}

\DeclareRobustCommand\sampleline[1]{%
  \tikz\draw[#1] (0,0) (0,\the\dimexpr\fontdimen22\textfont2\relax)
  -- (2em,\the\dimexpr\fontdimen22\textfont2\relax);%
}

\newcommand{\Rey}{\mathrm{Re}}
\newcommand{\Sto}{\mathrm{St}}
\DeclareRobustCommand\stl{\tikz[baseline]\draw[solid] (0,.5ex)--++(.5,0) ;}
\DeclareRobustCommand\dedash{\tikz[baseline]\draw[densely dashed] (0,.5ex)--++(.5,0) ;}
\DeclareRobustCommand\ddd{\tikz[baseline]\draw[dash dot dot] (0,.5ex)--++(.5,0) ;}
\DeclareRobustCommand\circles{\tikz \draw (0.5,0.1) circle (3pt);}
\DeclareRobustCommand\squares{\tikz \draw (0,0) rectangle (0.2,0.2);}
\DeclareRobustCommand\diamonds{\tikz \draw[rotate=45] (0,0) rectangle (0.2,0.2);}
\DeclareRobustCommand\triangles{\tikz \draw (0,0) -- (0.1,0.2) -- (0.2,0) -- cycle;}

\colorlet{revision}{black}

\begin{document}

\title{Mechanisms of drag reduction by semi-dilute inertial particles in turbulent channel flow}

\author{Himanshu Dave}
\affiliation{
  School for Engineering of Matter, Transport and Energy, Arizona State University, Tempe, AZ 85281, USA
}

\author{M. Houssem Kasbaoui}
\email{houssem.kasbaoui@asu.edu}
\affiliation{
  School for Engineering of Matter, Transport and Energy, Arizona State University, Tempe, AZ 85281, USA
}
\date{\today}

\begin{abstract}
We investigate the mechanisms by which inertial particles dispersed at semi-dilute conditions cause significant drag-reduction in a turbulent channel flow at $\Rey_\tau = 180$. We consider a series of four-way coupled Euler-Lagrange simulations where particles having friction Stokes number $\Sto^+ = 6$ or 30 are introduced at progressively increasing mass loading from $M=0.2$ to 1.0. The simulations show that $\Sto^+ = 30$ particles cause large drag-reduction by up to 19.74\% at $M=1.0$, whereas $\Sto^+ = 6$ particles cause large drag increase by up to 16.92\% at $M=1.0$. To reveal the mechanisms underpinning drag-reduction or drag-increase, we investigate the stress distribution within the channel and the impact of the dispersed particles on the near-wall coherent structures. We find a distinctive feature of drag-reducing particles which consists in the formation of extremely long clusters, called ropes. These structures align preferentially with the low-speed streaks and contribute to their stabilization and suppression of bursting. Despite the additional stresses due to the particles, the modulation of the near-wall coherent structures leads to a greater reduction of Reynolds shear stresses and partial relaminarization of the near-wall flow. In the case of the drag-increasing particles with $\Sto^+ = 6$, a reduction in Reynolds shear stresses is also observed, however, this reduction is insufficient to overcome the additional particle stresses which leads to drag increase.
\end{abstract}

\pacs{}
\maketitle

\section{Introduction}
\label{sec:introduction}

Inertial particles introduced in wall-bounded turbulent flows play a significant role in the transport of mass and momentum in many engineering applications. Examples include cyclone separators, fluidized bed risers, sediment transport in pipelines, and dust ingested in engines. In the case of gas-solid flows, \textcolor{revision}{semi-dilute particle concentrations, i.e., particle volume-fraction typically $10^{-6} - 10^{-3}$,  may be sufficient to cause significant modulation of the flow structures \citep{kasbaouiTurbulenceModulationSettling2019}. Provided that the mass loading is $O(1)$, the dynamics of the two phases in the semi-dilute regime are controlled by the two-way coupling between the particles and fluid, whereas, particle-particle collisions play a secondary or negligible role.} In this paper, we show that \textcolor{revision}{semi-dilute} inertial particles introduced in a turbulent channel flow may cause significant skin-friction drag increase  or reduction, depending on particle concentration, inertia, and how particle clusters interact with near-wall coherent flow structures.

Inertial solid particles, or liquid droplets small enough to be dominated by surface tension, dispersed in gas may drastically alter the carrier flow properties. Due to their inability to follow fluid streamlines, these particles exert micro-stresses on the carrier fluid. If the particles are sufficiently concentrated, the collective action of these micro-stresses may amount to a large macroscopic force capable of modifying the carrier flow properties.
Taking Homogeneously Sheared Turbulence (HST) as a simplified proxy for general turbulent shear flows, \citet{kasbaouiTurbulenceModulationSettling2019} and \citet{kasbaouiRapidDistortionTwoway2019} showed that introducing inertial particles at semi-dilute concentration, characterized by an average particle volume fraction $\phi_0 = 10^{-6} - 10^{-3}$ and mass loading $M=O(1)$, may cause an attenuation or augmentation of the turbulent kinetic energy. Whether the latter is increased or decreased depends on particle inertia. \citet{ahmedDirectNumericalSimulation2001} showed that turbulence modulation in HST is due to a reverse cascade of energy, whereby energy injected into the flow by particles at their scale propagates up to the macro-scales. The resulting particle-laden turbulence has distinctively different characteristics from turbulence in single-phase flows as shown by \citet{gualtieriClusteringTurbulenceModulation2013} who found that the energy cascade in particle-laden HST departs from the traditional -5/3 law \citep{kolmogorovLocalStructureTurbulence1941}.

In  wall-bounded turbulent flows, inertial particles are expected to have disproportionally larger impact on near-wall flow structures. Inertial particles tend to migrate to regions of lower turbulent fluctuations, a process called turbophoresis. The latter leads to the accumulation of particles near bounding walls. In simulations of dilute particle-laden turbulent channel flows, \citet{sardinaWallAccumulationSpatial2012}, and later, \citet{nilsenVoronoiAnalysisPreferential2013} and \citet{yuanThreedimensionalVoronoiAnalysis2018}, showed that the particle concentration in the viscous layer may be one or  two orders of magnitude larger than the mean. The highest wall accumulation happens for particles with inertia characterized by  friction Stokes number $\mathrm{St}^+=\tau_p u_\tau^2/\nu$ in the range 10-50. Here, $u_\tau$, $\nu$, and $\tau_p$ refer to the friction velocity, kinematic viscosity, and particle response time. Further, such inertial particles disperse in a highly inhomogeneous way leading to the formation of clusters even in the near-wall region where most particles accumulate \citep{sardinaWallAccumulationSpatial2012}. Clusters found therein tend to be elongated structures that may be several orders of magnitude longer than the particle diameter \citep{jieExistenceFormationMultiscale2022}.
Because inertial particles accumulate into such long clusters, they are able to modulate flow structures on scales as large as the cluster scales, which may exceed even the largest turbulence scales \citep{kasbaouiTurbulenceModulationSettling2019}. Hence, provided that the particle concentration is sufficiently large to yield meaningful feedback force on the flow, the dispersed particles are expected to modulate near-wall flow structures and alter the turbulence structure in wall-bounded flows.

The near-wall coherent flow structures have a large impact on skin-friction drag. The existence of these structures, their evolutionary dynamics and their role in the generation of shear stress in particle-free wall-bounded turbulent flows \textcolor{revision}{have long been under study.} \citet{fiedlerCoherentStructuresTurbulent1988} describes the existence of these structures within the boundary layer as a ``zoo of structures'' ranging from ``horseshoe- and hairpin-eddies, pancake- and surfboard-eddies, typical eddies, vortex rings, mushroom-eddies, arrowhead-eddies, etc''. In turbulent channel flows, \citet{jeongCoherentStructuresWall1997} found that flow structures in the buffer region are dominated primarily by elongated quasi-streamwise vortices. The latter are arranged antisymmetrically with vortices of opposite directions arranged next to each other \citep{schoppaCoherentStructureGeneration2002a,stretchAutomatedPatternEduction1991}. The so-called low-speed streaks are regions of slow moving fluid that have been identified in various studies, and were later shown to be nested in the space between a pair of quasi-streamwise vortices \citep{klineStructureTurbulentBoundary1967a,smithCharacteristicsLowspeedStreaks1983,jiangExperimentalStudyLowspeed2020}. The spanwise spacing of low-speed streaks  is a characteristic of turbulence in channels, since its value of $\sim$100 wall units was found to vary little with Reynolds number \citep{klineStructureTurbulentBoundary1967,jimenezMinimalFlowUnit1991,klewickiViscousSublayerFlow1995,jimenezLargescaleDynamicsNearwall2004}. Bursting occurs when the quasi-streamwise vortices become unstable \citep{jimenezCoherentStructuresWallbounded2018}. The formation and breakdown of these structures is part of a self-sustaining process that repeats periodically. \citet{willmarthStructureReynoldsStress1972} showed that bursting events are among the largest contributors to the Reynolds stress production. Thus, reducing skin-friction drag hinges on the ability to suppress bursting and stabilizing quasi-streamwise vortices as has been shown in drag-reduced polymeric channel flows \citep{mccombDragreducingPolymersTurbulent1978,bermanDragReductionPolymers1978,renardyMechanismDragReduction1995,zhuTransientDynamicsTurbulence2019}.

To the best of our knowledge, parameters leading to reproducible skin-friction drag reduction using inertial particles have not yet been identified. The majority of older studies point to an increase of skin-friction drag or negligible effect \citep{radinDragReductionSolidfluid1975a,gyrDragReductionTurbulence1995a}.
\citet{liNumericalSimulationParticleladen2001} are among the first to provide evidence of skin-friction drag reduction in simulations with the point-particle method. The authors showed that particles with friction Stokes number $\mathrm{St}^+=192$ dispersed in a vertical channel at $\Rey_\tau=u_\tau h/\nu = 125$, where $h$ is the channel half-height, increase the fluid mass flow rate by $\sim 5\%$ for mass loadings as small as $M=0.2$. Note that an increase in fluid mass flow rate is equivalent to a reduction in skin-friction drag. Later, \citet{zhaoTurbulenceModulationDrag2010} showed that inertial particles with $\mathrm{St}^+= 30$ at mass loading $M=0.36$ increase the fluid mass flow rate by approximately 15\% in a turbulent channel flow at  $\Rey_\tau=180$. However, these results may not be representative of a stationary state, since the latter requires much longer integration time than what is reported by \citet{zhaoTurbulenceModulationDrag2010}.  A follow-up study by \citet{zhouNonmonotonicEffectMass2020} in an identical configuration shows negligible drag reduction, about $\sim 0.2\%$ at $M=0.4$ and 2.8\% at $M=0.75$. Recently, \citet{costaNearwallTurbulenceModulation2021} revisited the semi-dilute particle-laden channel flow at $\Rey_\tau=180$ using particle-resolved direct numerical simulations (PR-DNS). Contrary to the aforementioned work,  \citet{costaNearwallTurbulenceModulation2021} found that inertial particles with $\mathrm{St}^+=50$ at $M=0.34$ cause a large increase in skin-friction drag by about $\sim 16\%$ compared to a particle-free channel. One should also note that despite the higher numerical resolution offered by PR-DNS, the greater computational cost constrained \citet{costaNearwallTurbulenceModulation2021} to use significantly smaller computational domain. With volume about 1/4th of that used in prior simulations with the point-particle method \citep{zhaoTurbulenceModulationDrag2010,zhouNonmonotonicEffectMass2020}, the computational box used by  \citet{costaNearwallTurbulenceModulation2021} may be too small to allow a natural development of particle clusters and their interaction with near-wall coherent structures.

Given the conflicting results previously reported, the questions of \emph{whether} inertial particles can induce significant skin-friction drag reduction, and if they do, \emph{how}?, have not been settled yet. 
\textcolor{revision}{
In this paper, we address these questions using Euler-Lagrange simulations of particle-laden turbulent channel flow at $\Rey_\tau=180$ while varying the characteristics of the particle phase. Although there is a multitude of non-dimensional numbers that can be used to characterize the particle-fluid interaction \citep{tanakaClassificationTurbulenceModification2008}, our past work shows that Reynolds number, Stokes number, and Mass loading are the most relevant non-dimensional numbers that control the dynamics in the semi-dilute regime \citep{kasbaouiPreferentialConcentrationDriven2015a,kasbaouiClusteringEulerEuler2019,kasbaouiRapidDistortionTwoway2019,kasbaouiTurbulenceModulationSettling2019,shuaiAcceleratedDecayLamb2022,shuaiInstabilityDustyVortex2022}. For this reason, we focus on varying the Stokes number $\Sto^+$ and mass loading $M$ seperately.}
 In section \ref{sec:setup}, we describe the mathematical framework, numerical methods, and simulation parameters used in this study. In order to highlight the flow modulation induced by inertial particles, we provide a brief review of the characteristics of particle-free turbulent channel flow at $\Rey_\tau = 180$ in section \ref{sec:unladen}, namely, in terms of velocity statistics, stress balance, and coherent flow structures. In section \ref{sec:particle_laden}, we analyze the particle-laden channel flows, in particular, the induced flow modification (\S~\ref{sec:modulation}), stress balance (\S~\ref{sec:stress_two_phase}), and interplay between particle clusters and near-wall coherent flow structures (\S~\ref{sec:mechanism}). Finally, we provide concluding remarks in \S~\ref{sec:conclusion}.

\section{Simulation setup and parameters}
\label{sec:setup}
In this section, we present the parameters and methods used in our Euler-Lagrange simulations of semi-dilute particle-laden turbulent channel flow. Section \ref{sec:formulation} provides an overview of the mathematical framework, while section \ref{sec:configuration} provides details about the configuration and flow parameters in this study.
\subsection{Mathematical formulation}
\label{sec:formulation}

The particle phase is treated in a Lagrangian frame, where each individual particle is tracked. For a particle ``$i$'', the dynamics are given by \citep{maxeyEquationMotionSmall1983}
\begin{eqnarray}
    \frac{d \bm{x}_p^i}{dt}&=&\bm{u}_p^i, \label{eq:MR_1}\\
    \frac{d \bm{u}_p^i}{dt}&=&f_d\frac{(\bm{u}_f(\bm{x}_p^i,t)-\bm{u}_p^i)}{\tau_p}+\frac{1}{\rho_p}\nabla\cdot\bm{\tau}(\bm{x}_p^i,t)+\bm{F}^i_{p\rightarrow p}+\bm{F}^i_{w \rightarrow p},\label{eq:MR_2}
\end{eqnarray}
where $\rho_p$, $d_p$, $\tau_p=\rho_p d_p^2/(18\mu)$, $\bm{x}_p^i$, $\bm{u}_p^i$, $\bm{F}^i_{w \rightarrow p}$, and $\bm{F}^i_{p\rightarrow p}$ are the particle density, diameter, response time, position, velocity, particle-wall collisional acceleration, and particle-particle collisional acceleration, respectively. \textcolor{revision}{The fluid stress tensor $\bm{\tau}$ is given by
\begin{equation}
\bm{\tau} = -p\bm{I}+\mu[\nabla\bm{u}_f + \nabla\bm{u}^T_f - \frac{2}{3}(\nabla\cdot\bm{u}_f)\bm{I}],\label{eq:tau_f}
\end{equation}
where the hydrodynamic $\bm{u}_f$ is the fluid velocity, $p$ is pressure, and $\mu$ is the dynamic viscosity. $\bm{I}$ is the identity tensor.} The first term on the right-hand side of (\ref{eq:MR_2}) accounts for the drag force exerted by the fluid on the particle. In order to accurately capture this force for particles with finite Reynolds number $\Rey_p=|\bm{u}_f-\bm{u}_p|d_p/\nu$ and particles that may be located within clusters, we use a nonlinear drag correction factor $f_d$ modeled with  the correlation by \citet{tennetiDragLawMonodisperse2011a}. The latter accounts for particle Reynolds number $\Rey_p$ and local particle volume fraction $\phi$. The second term on the right-hand side of (\ref{eq:MR_2}) represents the acceleration due to resolved fluid stresses on the particle phase. Although this term is included for completeness, its effect is negligible in the semi-dilute regime due to the high density ratio $\rho_p/\rho_f=O(10^3)$. For the same reason, other hydrodynamic forces are ignored. Note that particle-particle collisions, while typically negligible in dilute flows with average particle volume fraction $\phi_0<10^{-3}$, are included due to the tendency of inertial particles to accumulate into clusters with local volume fraction as high as $\phi\sim 10^{-2}$. \textcolor{revision}{The particle-particle and wall-particle collisions are performed using the soft-sphere collision model described in \citep{capecelatroEulerLagrangeStrategy2013}, and originally proposed \citet{cundallDiscreteNumericalModel1979}, with a restitution coefficient $e=0.9$. The unperturbed fluid velocity at the particle location is computed using the method of \citet{irelandImprovingParticleDrag2017}.} Further, in order to isolate inertial effects, gravity is ignored.
The carrier turbulent flow is described using volume-filtered incompressible Navier-Stokes equations \citep{andersonFluidMechanicalDescription1967,capecelatroEulerLagrangeStrategy2013},
\begin{eqnarray}
\frac{\partial }{\partial t}\left(1-\phi\right) + \nabla\cdot\left((1-\phi)\bm{u}_f\right)&=&0, \label{eq:NVS_1}\\
\rho_f\left( \frac{\partial }{\partial t}\left(\left(1-\phi\right) \bm{u}_f\right)+ \nabla \cdot\left(\left(1-\phi\right)\bm{u}_f\bm{u}_f\right)\right)&=&-\nabla p + \mu\nabla^2\bm{u}_f+(1-\phi)\bm{A}+\bm{F}_p+\nabla\cdot\bm{R}_\mu \label{eq:NVS_2},
\end{eqnarray}
where $\bm{u}_f$ is the fluid velocity, $p$ is pressure, $\bm{F}_p$ is the momentum exchange between the two-phases, and $\bm{A}$ is a constant pressure gradient that drives the flow within the channel. \textcolor{revision}{This forcing is a function of the wall shear stress $\tau_w$ and the channel half height $h$, such that $\bm{A} = (\tau_{w}/h)\bm{e}_x$, where $\bm{e}_x$ is a unitary vector oriented in the streamwise direction.} The tensor $\bm{R}_\mu$ arises from filtering the fluid stress tensor \citep{capecelatroEulerLagrangeStrategy2013}, and is closed using the effective viscosity model $\bm{R}_\mu=\mu_f((1-\phi)^{-2.8}-1)[\bm{\nabla}\bm{u}_f+\bm{\nabla}\bm{u}_f^T-\frac{2}{3}(\bm{\nabla}\cdot\bm{u}_f)I]$ of \citet{gibilaroApparentViscosityFluidized2007}. This term, included here for completeness, is negligible in the semi-dilute regime considered. Likewise, the presence of the volume fraction $\phi$ in equations (\ref{eq:NVS_1}) and (\ref{eq:NVS_2}) accounts for volume excluded by the particle phase \citep{andersonFluidMechanicalDescription1967}.  \textcolor{revision}{This effect is typically neglected in the semi-dilute regime \citep{paksereshtVolumetricDisplacementEffects2019}. However, we retain the volume fraction $\phi$ in our equations since turbophoresis and preferential concentration may lead to local accumulation of the particles resulting in volume fractions one or two orders of magnitude higher than the average \citep{sardinaWallAccumulationSpatial2012,nilsenVoronoiAnalysisPreferential2013,yuanThreedimensionalVoronoiAnalysis2018}.}
Consistently with equation (\ref{eq:MR_2}), the particles exert a feedback force on the fluid given by
\begin{equation}
    \bm{F}_p=-\phi \rho_p f_d \frac{\bm{u}_{f}\vert_p-\bm{u}_p}{\tau_p}-\phi \nabla\cdot\bm{\tau}\vert_p,\label{eq:Fp}
\end{equation}
where $\bm{u}_p(\bm{x},t)$ is the Eulerian particle velocity at the location $\bm{x}$, $\bm{u}_{f}\vert_p$ is the fluid velocity at the particle location, and $\bm{\tau}\vert_p$ is the total fluid stresses at the particle location. These Eulerian fields are computed from the Lagrangian quantities in (\ref{eq:MR_1}) and (\ref{eq:MR_2}) using a filtering procedure that reads
\begin{eqnarray}
    \phi(\bm{x},t)&=&\sum_{i}^N V_{p}q\left(\left|\left|\bm{x}-\bm{x}_p^i\right|\right|\right),\\
    \phi\bm{u}_{p}(\bm{x},t)(\bm{x},t)&=&\sum_{i}^N V_{p}\bm{u}_p^iq\left(\left|\left|\bm{x}-\bm{x}_p^i\right|\right|\right),\\
    \phi\bm{u}_{f}\vert_p(\bm{x},t)&=&\sum_{i}^N V_p\bm{u}_f (\bm{x}_p^i (t),t) q\left(\left|\left|\bm{x}-\bm{x}_p^i\right|\right|\right),
\end{eqnarray}
where $V_p=\pi d_p^3/6$ is the particle volume and $q$ is a Gaussian filter kernel. As discussed above, drag force dominates the momentum exchange in the semi-dilute regime. From a scaling analysis of equation (\ref{eq:Fp}), one can see that the particle feedback force scales with mass loading $M=\phi_0\rho_p/\rho_f$. Consequently, the feedback force from the particle phase onto the fluid phase is negligible if $M\ll 1$. In this case, the flow dynamics are independent from those of the particle phase, essentially behaving as in particle-free conditions. However, as $M$ approaches unity, the coupling between the two-phases strengthens resulting in flow and particle dynamics that are mutually interlinked. The dynamics resulting from the joint evolution of the particle and fluid phases at $M=O(1)$ are the subject of this study.

\subsection{Channel flow configuration}
\label{sec:configuration}
\begin{table}
\caption{Summary of the non-dimensional parameters.\label{tab:parameters}}
\begin{ruledtabular}
\begin{tabular}{l l l l l l l l l}
 Runs & $\Rey\tau$ & $\Sto^+$ & $M$ & $\phi_0$        & $d_p^+$ & $\rho_p/\rho_f$ & $h/d_p$ & $N_p$\\ \hline
 A    & 180        & 6        & 0.2 & $2.4\times 10^{-4}$ & 0.36    & 833             & $500$   & $6.03 \times 10^6$ \\
 B    & 180        & 6        & 0.6 & $7.2\times 10^{-4}$ & 0.36    & 833             & $500$   & $18.1 \times 10^6$ \\
 C    & 180        & 6        & 1.0 & $1.2\times 10^{-3}$ & 0.36    & 833             & $500$   & $30.1 \times 10^6$ \\
 D    & 180        & 30       & 0.2 & $2.4\times 10^{-4}$ & 0.80    & 833             & $225$   & $4.93 \times 10^6$ \\
 E    & 180        & 30       & 0.6 & $7.2\times 10^{-4}$ & 0.80    & 833             & $225$   & $14.8 \times 10^6$ \\
 F    & 180        & 30       & 1.0 & $1.2\times 10^{-3}$ & 0.80    & 833             & $225$   & $24.7 \times 10^6$ \\
\end{tabular}
\end{ruledtabular}
\end{table}

\begin{figure}
  \begin{subfigure}{0.4\textwidth}
    \centering
    \includegraphics[width=1\linewidth]{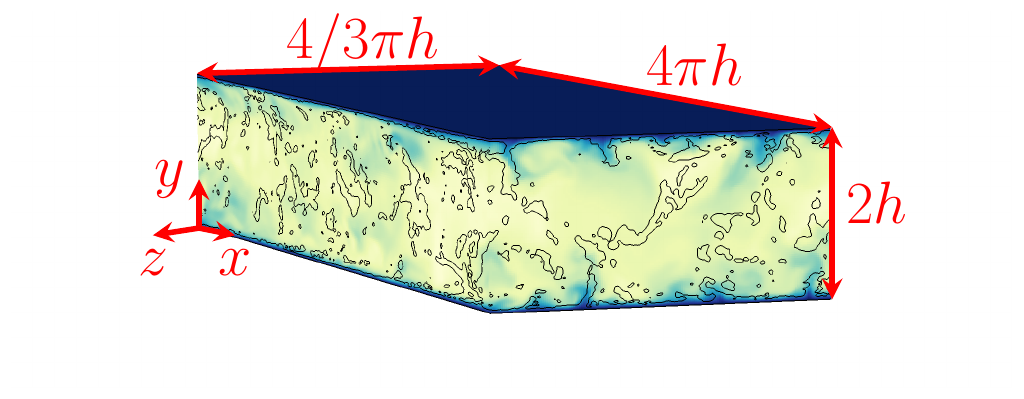}
    \caption{}
    \label{fig:configuration_s}
  \end{subfigure}
  \begin{subfigure}{0.5\textwidth}
    \centering
    \includegraphics[width=1\linewidth]{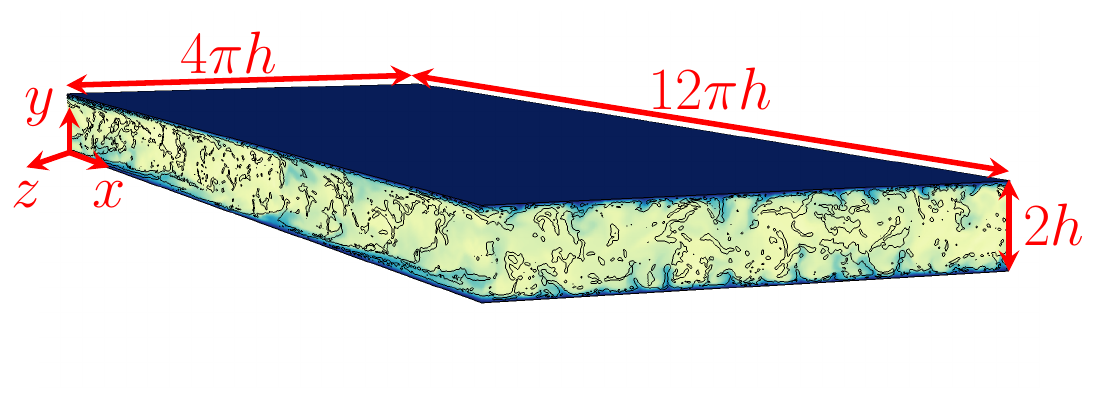}
    \caption{}
    \label{fig:configuration_l}
    \end{subfigure}
    \caption{Schematic of the computational domains used in simulations of channels laden with (a) $\Sto^+=6$ and (b) $\Sto^+=30$ particles.}
    \label{fig:configuration}
\end{figure}

We consider 6 mono-disperse particle-laden turbulent channel flows at varying particle-phase properties. Table \ref{tab:parameters} provides a summary of the flow and simulation parameters. In all these simulations, the friction Reynolds number is fixed at $\Rey_\tau=\rho_f u_\tau h/\mu=180$, where $u_\tau=\sqrt{\tau_w/\rho_f}$ is the friction velocity and $\tau_w$ is the wall shear stress.  Two particle diameters are considered yielding non-dimensional diameters $d_p^+=0.36$ and $0.80$. The superscript $+$ denotes a quantity that has been non-dimensionalized using inner wall scaling.  These particles are sufficiently small to make any finite-size effects negligible. The friction Stokes number, which measures particle inertia, is $\Sto^+=\tau_pu_\tau^2/\nu=6$ and $\Sto^+=30$ for the particles with $d_p^+=0.36$ and $d_p^+=0.80$, respectively. In both cases, particle inertia is significant such that one may expect these particles to form clusters and accumulate near walls due to turbophoresis \citep{sardinaWallAccumulationSpatial2012}. For each of the two Stokes numbers considered, we vary the average particle volume fraction to yield $\phi_0 = 2.4\times 10^{-4}, 7.2\times 10^{-4}$, and $1.2\times 10^{-3}$. With the particle-fluid density ratio fixed at $\rho_p/\rho_f=833$ for all 6 cases, the mass loading $M=\rho_p/\rho_f\phi_0$ is 0.2, 0.6, or 1.0. These parameters correspond to the semi-dilute regime, where the particle phase is dilute, yet the two-way coupling between the particle and fluid phases is strong due to the large mass loading. Thus, particle feedback on the fluid can be expected to lead to significant modulation of the flow, especially for cases with $M=0.6$ and $1.0$.

Figure \ref{fig:configuration} shows a schematic of the domains used in this study. Channel flows laden with  $\Sto^+=6$ particles are simulated in a domain of size $4\pi h$ in the streamwise direction $x$, $2h$ in the wall-normal direction $y$, and $(4/3)\pi h$ in the spanwise direction $z$ as shown in figure \ref{fig:configuration_s}. This domain size is comparable to those used in previous studies \citep{zhaoTurbulenceModulationDrag2010,zhouNonmonotonicEffectMass2020}. Channel flows laden with  $\Sto^+=30$ particles are simulated in a domain 9 times larger with dimensions $12\pi h \times 2h \times 4\pi h$ as shown in figure \ref{fig:configuration_l}. \textcolor{revision}{While computations in such larger domain are significantly more expensive, we have found it necessary to use this larger domain to capture the increased spanwise spacing of the particle and flow structures with $\Sto^+=30$ particles. This aspect is discussed in \S\ref{sec:mechanism}.}

The equations of motion are solved using the flow solver NGA \citep{desjardinsHighOrderConservative2008}, with the Euler-Lagrange strategy of \citet{capecelatroEulerLagrangeStrategy2013}. The fluid mass and momentum equations (\ref{eq:NVS_1}) and (\ref{eq:NVS_2}) are solved on a staggered grid of size $226\times128\times 168$ for the small channel and \textcolor{revision}{$678\times128\times 504$} for the larger one. The grid is stretched in the wall-normal direction using a hyperbolic tangent function such that the minimum mesh spacing in the wall-normal direction is $\Delta y^+_\mathrm{min}=0.5$. In the streamwise and spanwise directions, the mesh spacing is constant at $\Delta x^+=10$ and $\Delta z^+=5$, respectively. In both small and large domains, the values $\Delta x^+$, $\Delta y^+_\mathrm{min}$, and $\Delta z^+$ are identical. The discretization relies on second order finite-volume schemes that preserve mass, momentum and kinetic energy \citep{morinishiFullyConservativeFinite2004,morinishiFullyConservativeHigher1998,desjardinsHighOrderConservative2008}. The fluid equations are advanced in time with a fractional step approach and a Crank-Nicolson scheme introduced by \citet{pierceProgressvariableApproachLargeeddy2004}. Equations (\ref{eq:MR_1}) and (\ref{eq:MR_2}), describing the position and velocity of Lagrangian particle are advanced using a second-order Runge-Kutta scheme. \textcolor{revision}{Soft-sphere particle-particle and particle-wall collisions are handled with the method in \citep{capecelatroEulerLagrangeStrategy2013}.} Depending on the case, a total of \textcolor{revision}{$N=4.9\times 10^6$ to $30.1\times 10^6$} particles are tracked in the simulation domain. Eulerian particle data such as the volume fraction field are computed from Lagrangian data using a Gaussian kernel of width $7d_p$. The method is fully conservative, yields grid-independent solutions in two-way coupled problems, and has been extensively verified against experiments \citep{capecelatroMassLoadingEffects2015,capecelatroNumericalCharacterizationModeling2014,capecelatroEulerianLagrangianModeling2013,wangInertialParticleVelocity2019}, and theoretical calculations \citep{kasbaouiRapidDistortionTwoway2019,kasbaouiTurbulenceModulationSettling2019,shuaiAcceleratedDecayLamb2022,shuaiInstabilityDustyVortex2022}.

The Euler-Lagrange simulations are initialized from auxiliary simulation of unladen channel flow at $\Rey_\tau=180$. Once the single-phase flow reaches a stationary state, the Lagrangian particles are inserted randomly into the channel with velocities matching the fluid velocity interpolated at their locations. To reach a new stationary state, the two-phase flow simulations are carried out for 120 eddy turnover times ($h/u_\tau$). After which, the simulations are run for additional 10 eddy turnover times to collect statistics. \textcolor{revision}{In total, running these simulations required $2.58$M CPU hours (cpu.h) on Intel Xeon Gold 6252  nodes, with each $\Sto^+=6$ simulation requiring 345,600 cpu.h and each $\Sto^+=30$ simulation requiring 518,400 cpu.h.}

\section{Structure of a particle-free turbulent channel flow}
\label{sec:unladen}

\begin{figure}
  \begin{subfigure}{0.45\textwidth}
    \centering
    \includegraphics[width=1\linewidth]{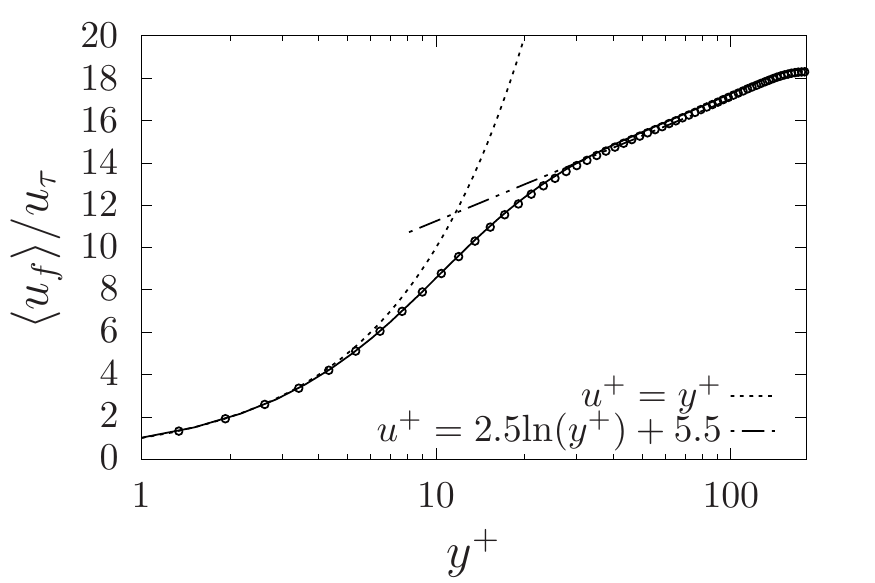}
    \caption{ }
    \label{fig:unladen_vel}
  \end{subfigure}
  \begin{subfigure}{0.45\textwidth}
    \centering
    \includegraphics[width=1\linewidth]{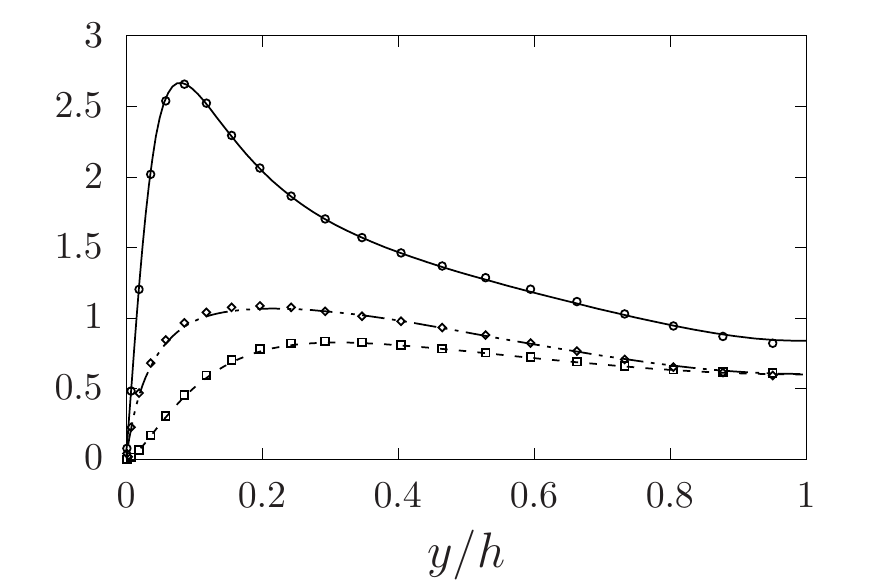}
    \caption{ }
    \label{fig:unladen_rms}
    \end{subfigure}
    \caption{Unladen flow velocity statistics at $\Rey_\tau =180$: (a) Streamwise mean velocity (b) root-mean square velocity fluctuations in the streamwise ($u_{f,rms}^+$, \stl), normal ($v_{f,rms}^+$, \dedash) and spanwise ($w_{f,rms}^+$, \ddd) directions. The symbols correspond to the data taken from 
    \citet{kimTurbulenceStatisticsFully1987}.}
    \label{fig:unladen_results}
\end{figure}

The particle-free channel represents a baseline reference that we use to highlight flow modifications induced by inertial particles. To that end, we start by reviewing aspects of an unladen turbulent channel flow at $\Rey_\tau=180$ that are relevant to the discussion in \S\ref{sec:particle_laden}.

Figure \ref{fig:unladen_results} shows  profiles of  mean streamwise velocity $\langle u^+_f\rangle$ and root-mean square (rms) velocity  fluctuations. Averaging is performed using 100 snapshots collected over a period of 10 of eddy turnover time once the flow is stationary. Further,  the streamwise and spanwise directions are averaged such that the only variation is in the wall-normal direction. As expected at this Reynolds number, the mean streamwise velocity shows three characteristic regions: viscous layer for $y^+\lesssim 5$, a buffer layer for $5\lesssim y^+\lesssim 30$, and a logarithmic layer for $y^+\gtrsim 30$. Velocity fluctuations in the streamwise direction dominate over the two other components and peak at $y^+\sim 12$ in the buffer layer. These observations are consistent with those of \citet{kimTurbulenceStatisticsFully1987} and general understanding of turbulent channel flow at the Reynolds number considered.

\begin{figure}
    \centering
    \includegraphics[width=5.0in]{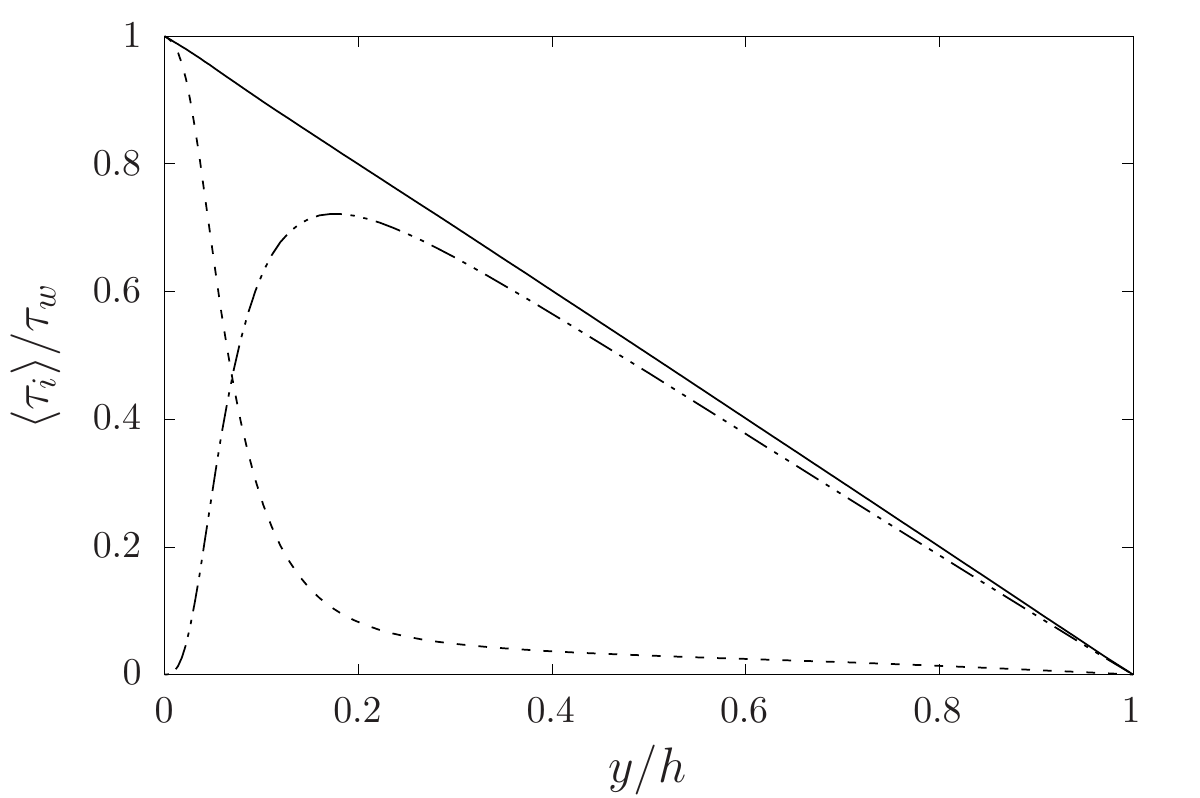}
    \caption{Contribution of the Reynolds shear stress(\ddd) and viscous shear stress(\dedash) to the total shear stress(\stl) in the particle-free channel at $\Rey_\tau = 180$.}
    \label{fig:tss_sp}
\end{figure}

The structure of the mean flow results from a balance between pressure gradient $-\langle\partial p/\partial x\rangle$, viscous stress $\mu d\langle u_f\rangle/dy$, and Reynolds shear stress $-\rho_f \langle u'_fv'_f \rangle $. By Reynolds-averaging the fluid momentum equations, one can show that the equation for the streamwise momentum reduces to
\begin{equation}
    -\langle\frac{\partial  p }{\partial x}\rangle = \frac{d}{d y}\left(\mu\frac{d}{dy} \langle u_f\rangle - \rho_{f}\langle u'_{f}v'_{f}\rangle\right).
    \label{eq:RSS_1}
\end{equation}
Given that the pressure gradient in a fully developed channel is constant, the total shear stress (sum of the viscous and Reynolds stresses) must vary linearly across the channel, i.e.,
\begin{equation}
    \mu\frac{d}{dy} \langle u_f\rangle - \rho_{f}\langle u'_{f}v'_{f}\rangle = \tau_{w}\left(1-\frac{y}{h}\right).
    \label{eq:cf_rss}
\end{equation}
This behavior is illustrated in figure \ref{fig:tss_sp} showing the variation of the total shear stress and its viscous and Reynolds contributions as a function of the wall normal distance. In accordance with (\ref{eq:cf_rss}), the total shear stress varies linearly from the wall to the channel center where it vanishes. The viscous shear stress dominates near the wall, and vanishes away from it. Conversely, the contribution of the Reynolds shear stress is small in the viscous sublayer, whereas it dominates in the logarithmic layer.

The Reynolds shear stresses has a direct influence on skin-friction drag. The latter is characterized using the coefficient
\begin{equation}
    C_{f}=\frac{\tau_w}{\frac{1}{2}\rho_f U_{b,f}^{2}},
    \label{eq:Cf}
\end{equation}
where $U_{b,f}=\dot{m}_f/\rho_f$ is the bulk fluid velocity corresponding to the ratio of the cross-sectional average fluid mass flow rate and the fluid density.  While $\tau_w$ is fixed in a channel flow driven by a constant pressure gradient, modulating the Reynolds shear stress is susceptible to change the bulk velocity $U_{b,f}$, and, in turn, the skin-friction drag $C_f$. Double integrating equation (\ref{eq:cf_rss}) clarifies the connection between $C_f$ and Reynold shear stress. The resulting mass flow rate per unit spanwise length is
\begin{equation}
    \frac{\dot{m}_f}{L_z}= \frac{2}{3} \frac{\tau_w h^2}{\nu}\left(1 + \frac{3}{(u_\tau h)^2}\int_{y=0}^{h} \int_{y'=0}^y \langle u'_fv'_f\rangle dy'dy\right).
    \label{eq:sp_mass_flow_rate}
\end{equation}
In this form, it becomes clear that the Reynolds shear stress reduces the mass flow rate, given $\langle u'_fv'_f\rangle<0$, resulting in an increase of $C_f$ compared to the laminar baseline. Therefore, it is not surprising that a large number of prior studies on skin-friction drag reduction in turbulent channel flows focused on reducing the Reynolds shear stress \citep{hetsroniLowspeedStreaksDragreduced1998,minDragReductionPolymer2003,ptasinskiTurbulentChannelFlow2003}.

\begin{figure}
	\centering
  \includegraphics[width=\linewidth]{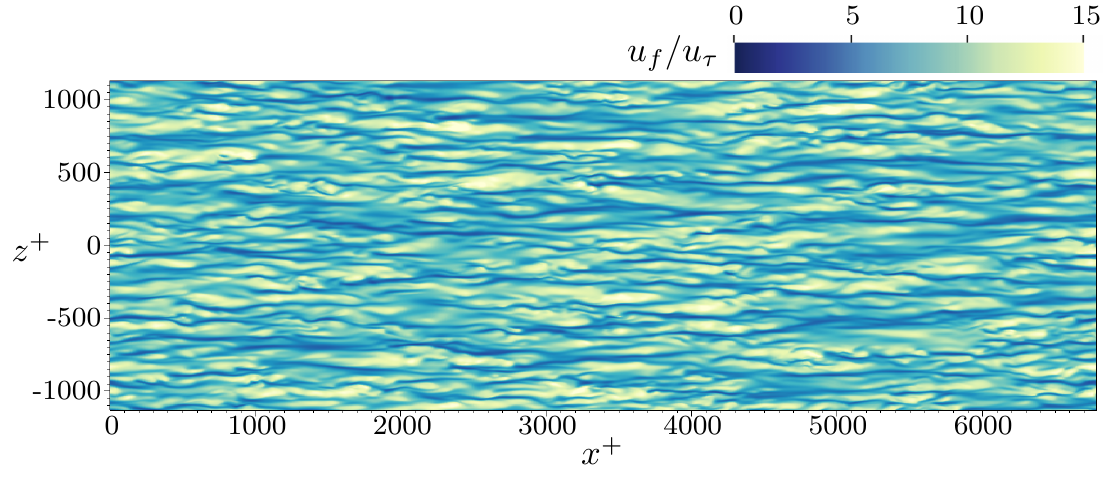}
	\caption{Isocontours showing the low speed streaks in the particle-free channel at $y^+ = 10$.
	\label{fig:streak_SP_a}}
\end{figure}

\begin{figure}
	\centering
  \includegraphics[width=5.0in]{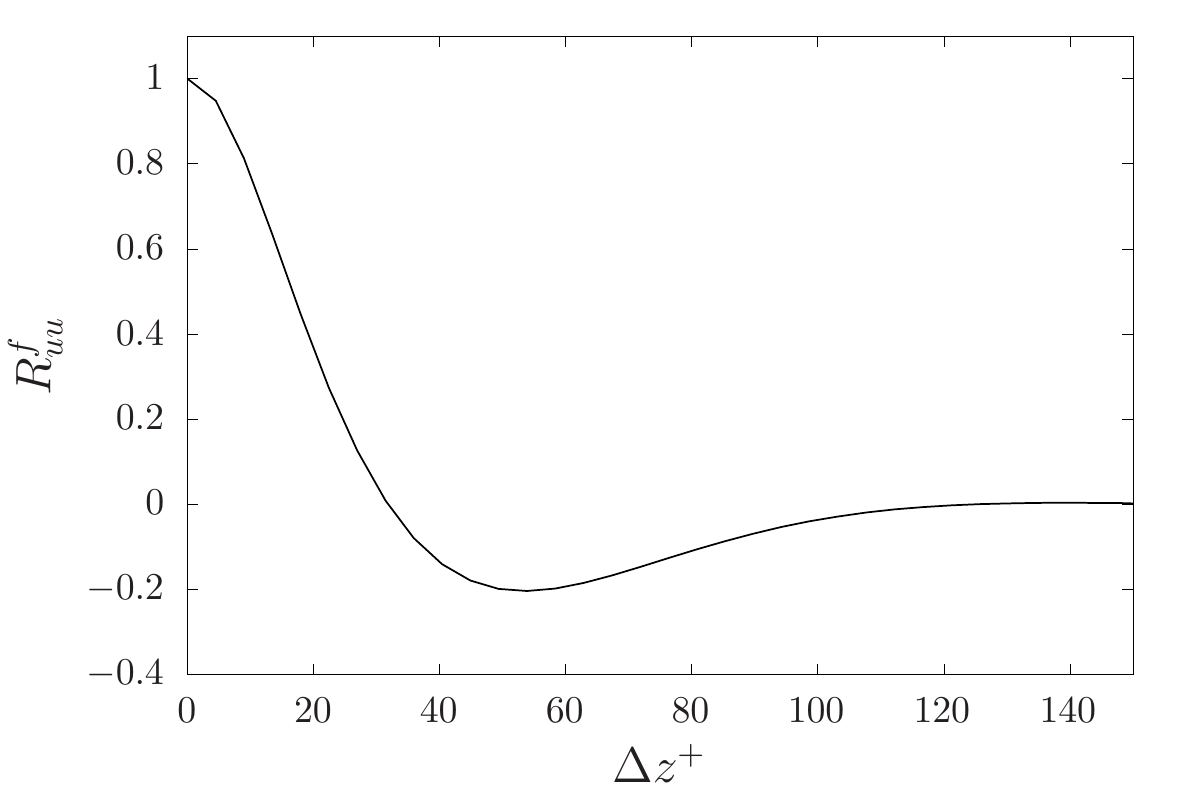}
	\caption{Two-point autocorrelation of the streamwise velocity fluctuations in the spanwise direction for particle-free channel at $y^+ = 10$.
	\label{fig:streak_SP_b}}
\end{figure}

From a mechanistic perspective, the Reynolds shear stress arises from coherent flow structures that populate the near-wall region \citep{smitsHighReynoldsNumber2011}. The so-called low-speed streaks, regions of slow moving fluid elongated in the streamwise direction, are among the most significant coherent structures found in a turbulent channel flow \citep{baeLifeCycleStreaks2021}. These streaks are shown in figure \ref{fig:streak_SP_a} visualized using isocontours of streamwise velocity in a wall parallel plane at $y^+=10$.
There has been sustained effort to understand the morphology and dynamics of low-speed streaks, as well as their connection to other coherent structures, such as quasi-streamiwse vorticies and so-called large-scale motions and very large-scale motions (see
\citep{klineStructureTurbulentBoundary1967a,willmarthStructureReynoldsStress1972,smithCharacteristicsLowspeedStreaks1983,jimenezMinimalFlowUnit1991,stretchAutomatedPatternEduction1991,jeongCoherentStructuresWall1997,jimenezCoherentStructuresWallbounded2018,zhuVortexAxisTracking2019,jiangExperimentalStudyLowspeed2020,schoppaCoherentStructureGeneration2002,smitsReynoldsStressScaling2021,baeLifeCycleStreaks2021,zhouInteractionNearwallStreaks2022}). The general consensus is that low-speed streaks in the buffer layer are formed between a pair of quasi-streamwise vortices with opposite orientation. The bursting of low-speed streaks contributes a significant part of the Reynolds shear stress and turbulent energy production \citep{kimProductionTurbulenceSmooth1971,baeLifeCycleStreaks2021}. This occurs when the quasi-streamwise vortices surrounding a low-speed streak become unstable \citep{jeongCoherentStructuresWall1997}, resulting in a lift up and eventual break down of the streak. Prior to their collapse, low-speed streaks in the buffer region have a typical length in the range 200--300 wall units \citep{baeLifeCycleStreaks2021,jeongCoherentStructuresWall1997}, but may extend beyond 1000 wall units \citep{jimenezLargescaleDynamicsNearwall2004}. Remarkably, these structures have a spanwise spacing $\lambda_f^+$ that varies little with Reynolds number, and is about $\lambda_f^+ \sim 100-110$ \citep{klineStructureTurbulentBoundary1967,jimenezMinimalFlowUnit1991,klewickiViscousSublayerFlow1995,jimenezLargescaleDynamicsNearwall2004}. We verify this by computing the two-point autocorrelation of the streamwise velocity as a function of the spanwise separation and wall distance,
\begin{equation}
    R^f_{uu}(\Delta z;y_0)=\frac{\langle{u'_f (x,y_0,z,t)u'_f (x,y_0,z+\Delta z,t)}\rangle}{\langle{u_f'^2}\rangle}.
    \label{eq:lambda_f}
\end{equation}
Figure \ref{fig:streak_SP_b} shows the variation of  $R^f_{uu}$ at $y^+=10$. The streak spanwise spacing $\lambda_f^+$ corresponds to twice the distance between the origin and $\Delta z^+$ where $R^f_{uu}$ reaches a first minimum which yields $\lambda_f^+ =2\times 53= 106$ in the present simulations.

It is noteworthy that the physics of a turbulent channel flow, discussed here at $\Rey_\tau =180$, remain largely the same near the wall, even at much higher Reynolds numbers. \citet{moserDirectNumericalSimulation1999} conduct wall-bounded channel flow simulations at $\Rey_\tau = 590,395$ and $180$. They showed that despite differences in the log region, the dynamics in the viscous and buffer layer regions are similar. Given that inertial particles tend to accumulate in these two regions, we expect that the particle-fluid interactions observed at $\Rey_\tau = 180$ will persist to much higher Reynolds numbers.

\section{Effect of inertial particles at semi-dilute conditions}
\label{sec:particle_laden}

Introducing inertial particles at semi-dilute concentration causes a departure from the known characteristics of a particle-free channel flow. In the following, we analyze the flow modulation resulting from the particle feedback force and propose a mechanism based on the interplay between near-wall coherent structures and particle clusters.

\subsection{Flow modulation and impact on skin-friction drag}
\label{sec:modulation}

\begin{figure}
	\centering
	\includegraphics[width=5in]{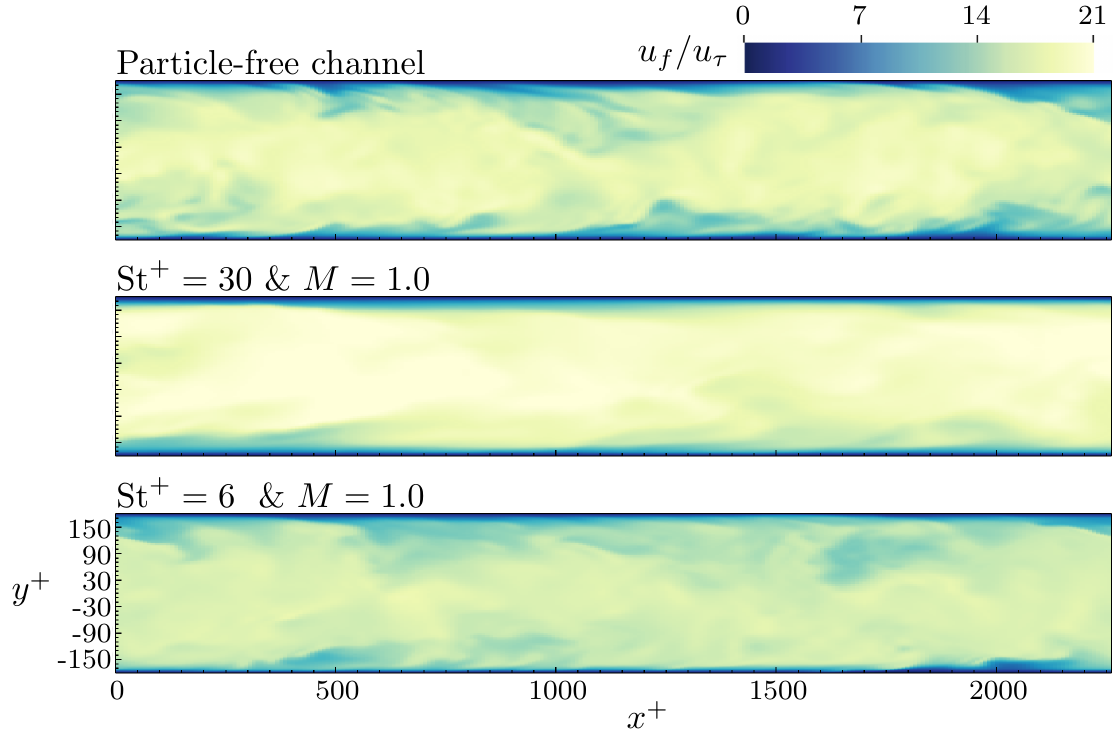}
	\caption{Isocontours of streamwise velocity in a wall-normal plane: (top) the reference unladen flow, (middle) flow laden with $\Sto^+=30$ particles at $M=1.0$, and (bottom) flow laden with $\Sto^+=6$ particles at $M=1.0$. Note: To facilitate visual comparison, the domain in the middle has been truncated to the same dimensions as the smaller domain in the bottom case.
	\label{fig:isocontour_u}}
\end{figure}

Figure \ref{fig:isocontour_u} shows isocontours of the streamwise velocity at an arbitrary time after the flow reached a stationary state.
From these instantaneous visualizations, it is immediately clear that semi-dilute inertial particles cause strong modulation of the carrier flow, with the most apparent change being a change in fluid bulk velocity. The latter is visibly increased by $\Sto^+=30$ particles at mass loading $M=1$ compared to the reference particle-free flow.
 In particular, the fluid streamwise velocity near the centerline shows a noticeable increase.
Further, the overall level of turbulence judged by the naked eye is diminished compared to the unladen flow.
In contrast, $\Sto^+=6$ particles at mass loading $M=1$ cause an apparent slow down of the carrier flow. The greatest drop of the fluid velocity is around the centerline.

\begin{figure}\centering
  \begin{subfigure}{0.49\linewidth}
    \centering
    \includegraphics[width=3.2in]{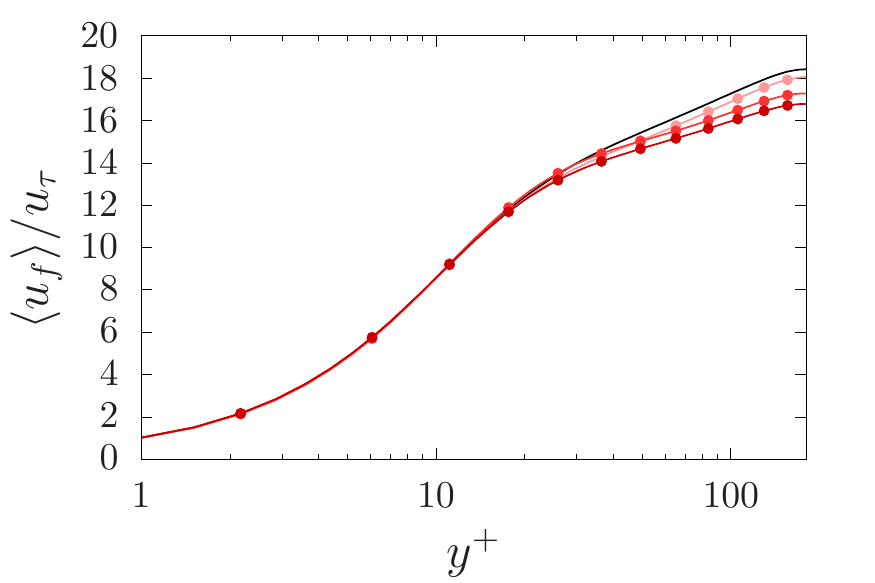}
    \caption{}
    \label{fig:u_st6}
  \end{subfigure}\hfill
  \begin{subfigure}{0.49\linewidth}
    \centering
    \includegraphics[width=3.2in]{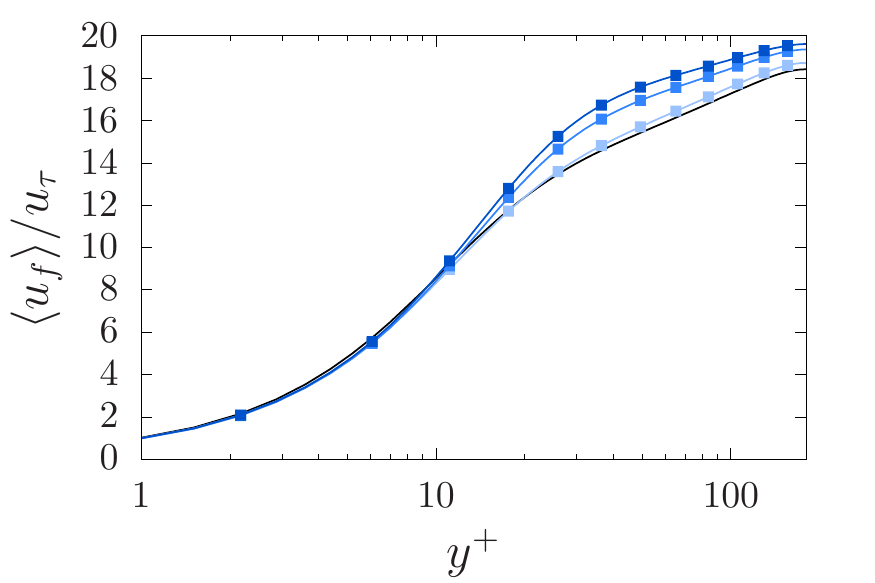}
    \caption{}
    \label{fig:u_st30}
	\end{subfigure}
  \caption{Average streamwise velocity as a function of the wall normal distance for (a) $\Sto^+ = 6$ and (b) $\Sto^+ = 30$ various mass loadings. The solid black line represents a particle-free channel. Symbols denote Stokes number $\Sto^+=$ 6 (\textcolor{red}{\protect\scalebox{1.75}{$\bullet$}}) or 30 (\textcolor{blue}{$\blacksquare$}). Darker symbols correspond to larger mass loading which varies from 0.2 to 1.0.}
  \label{fig:uavg}
\end{figure}

\begin{figure}\centering
  \begin{subfigure}{0.49\linewidth}
    \centering
    \includegraphics[width=3.20in]{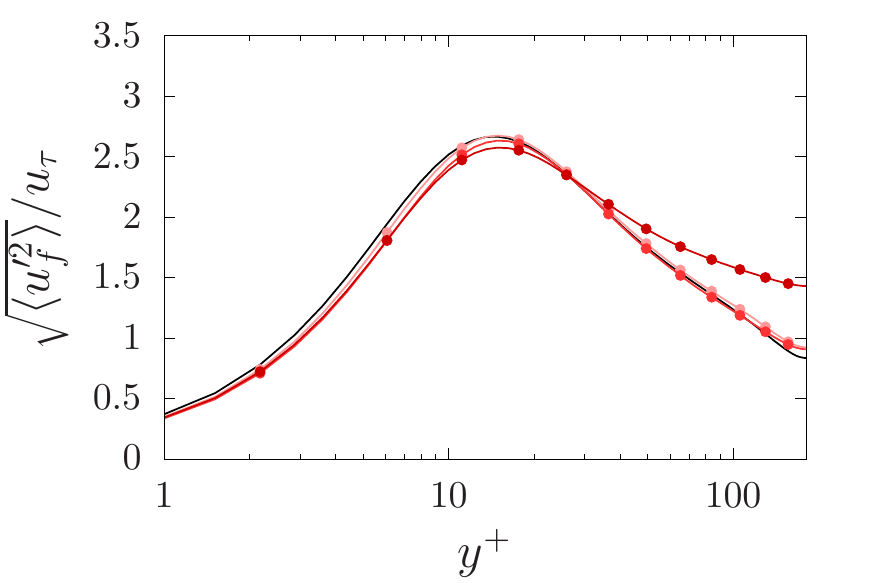}
    \caption{}
    \label{fig:u_rms_st6}
  \end{subfigure}\hfill
  \begin{subfigure}{0.49\linewidth}
    \centering
    \includegraphics[width=3.20in]{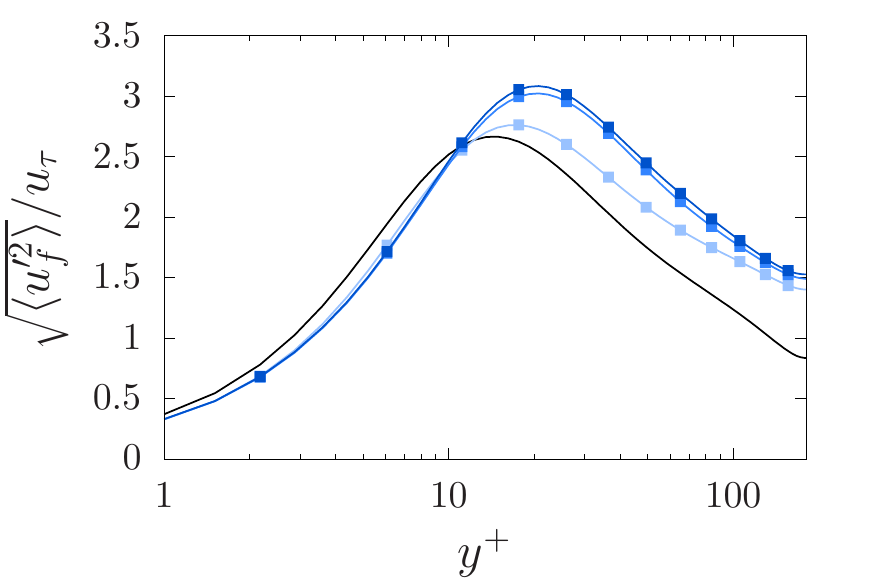}
    \caption{}
    \label{fig:u_rms_st30}
  \end{subfigure}\\
  \begin{subfigure}{0.49\linewidth}
    \centering
    \includegraphics[width=3.20in]{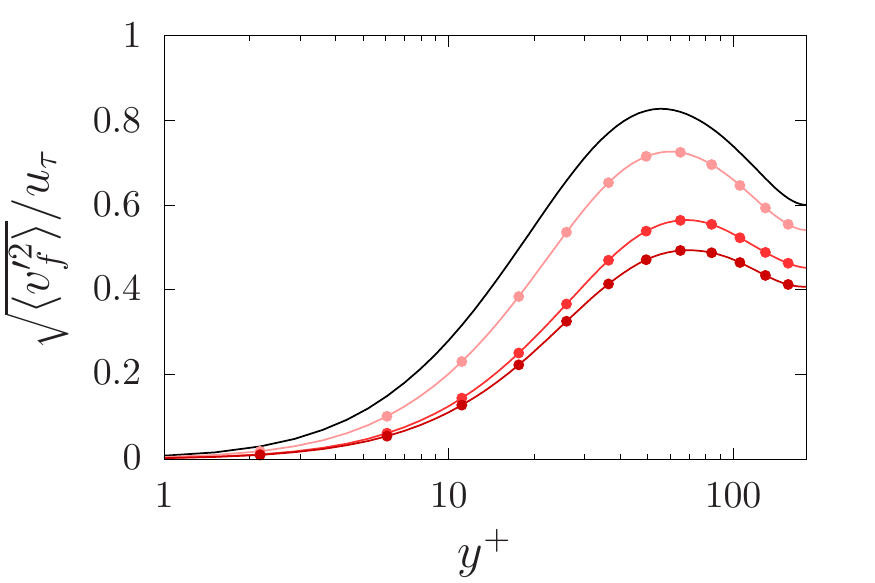}
    \caption{}
    \label{fig:v_rms_st6}
  \end{subfigure}\hfill
	\begin{subfigure}{0.49\linewidth}
		\centering
		\includegraphics[width=3.20in]{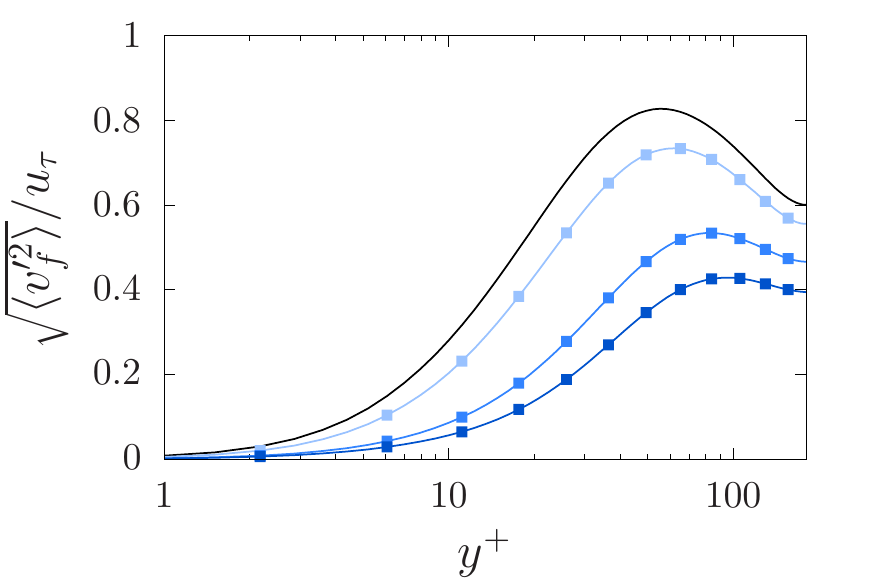}
		\caption{}
		\label{fig:v_rms_st30}
	\end{subfigure}\\
	\begin{subfigure}{0.49\linewidth}
		\centering
		\includegraphics[width=3.20in]{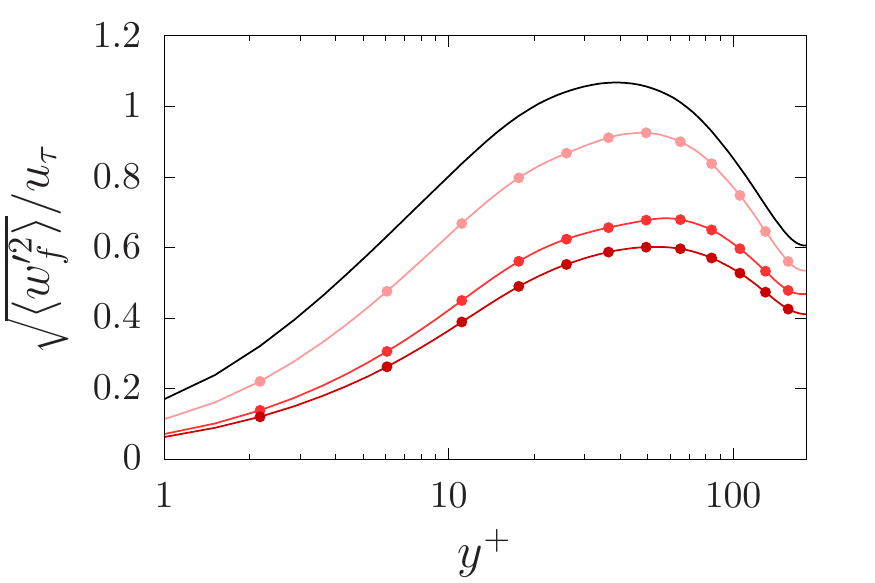}
		\caption{}
		\label{fig:w_rms_st6}
	\end{subfigure}\hfill
	\begin{subfigure}{0.49\linewidth}
		\centering
		\includegraphics[width=3.20in]{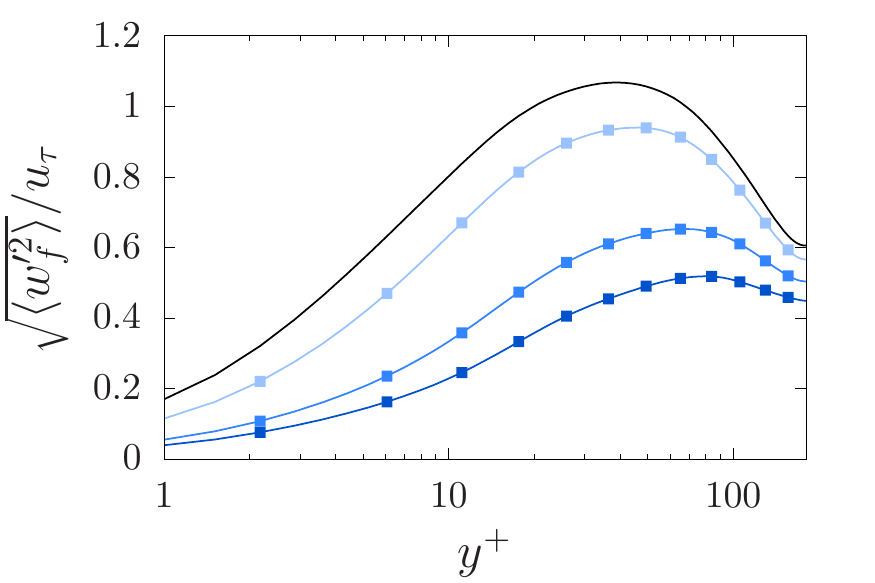}
		\caption{}
		\label{fig:w_rms_st30}
	\end{subfigure}
  \caption{Variation of the fluid root-mean-square velocity fluctuations with the wall normal distance for $\Sto^+ = 6$ (a,c,e) and $\Sto^+ = 30$ (b,d,f) at $M = 0.2-1.0$. The solid black line represents a particle-free channel. Symbols as in figure \ref{fig:uavg}.
  \label{fig:vel_rms}}
\end{figure}

Figure \ref{fig:uavg} shows the mean streamwise velocity profile for the particle-laden cases with $\Sto^+ = 6$ and $\Sto^+ = 30$ at the mass loadings $M=0.2$, 0.6, and 1.0. The velocity profile in the particle-free channel flow is also shown for comparison. In the viscous sublayer, the profiles follow the same linear scaling as the unladen channel. The largest impact of the inertial particles manifests in the buffer and logarithmic layers. In channels laden with $\Sto^+=30$ particles, the fluid velocity profile shifts upward in the logarithmic layer from the reference profile of the unladen channel. This trend is further reinforced with increasing mass loading which leads to greater upward shift of the velocity profile. These observations are in agreement with those of \citet{zhouNonmonotonicEffectMass2020} who found a similar upward shift of the streamwise velocity profile in their simulations at $\Rey_\tau=180$, $\Sto^+=30$, and $M=0.75$. Conversely, the profile of the mean streamwise velocity shifts downward from the reference unladen flow when $\Sto^+=6$ particles are suspended. Similar to the cases with higher inertia particles, increasing the mass loading causes an amplification of the trend observed, i.e., a downward shift of the profile here.

Figures \ref{fig:isocontour_u} and \ref{fig:uavg} provide qualitative and quantitative evidence that $\Sto^+=30$ particles increase the fluid mass flow rate, while $\Sto^+=6$ particles decrease it. As discussed in \S\ref{sec:unladen}, the implication of this flow modulation in a channel driven by a constant pressure gradient is that $\Sto^+=30$ particles decrease skin-friction drag, whereas $\Sto^+=6$ perform the opposite, i.e., increase skin-friction drag.

Figures \ref{fig:vel_rms} shows the variation of the velocity fluctuations root-mean-square (r.m.s.) with Stokes number and mass loading. For $\Sto^+=6$ particles, the profile of the streamwise fluctuations  at $M=0.2$ and $M=0.6$ changes little from the profile of the unladen flow. Only when mass loading is increased to $M = 1.0$ do we see a significant change of the streamwise fluctuations, primarily in the log region. Conversely, $\Sto^+=30$ particles lead to a more pronounced modulation of the streamwise velocity fluctuations at all three mass loadings considered. Generally, the streamwise fluctuations decrease slightly in the viscous layer and increase significantly in the buffer and logarithmic layers. Further, the location of the peak shifts from about $y^+=14$ in the unladen case to $y^+=22$ at $M=1.0$. While $\Sto^+=30$ and $\Sto^+=6$ particles exhibit different modulation characteristics for the fluid streamwise velocity fluctuations, their impact on the wall normal and spanwise velocity fluctuations displays fewer differences. Both particles cause significant dampening of the wall normal and spanwise fluctuations in the viscous, buffer and logarithmic layers. Increasing mass loading leads to larger reduction of these fluctuations.

\begin{figure}
  \includegraphics[width=5.0in]{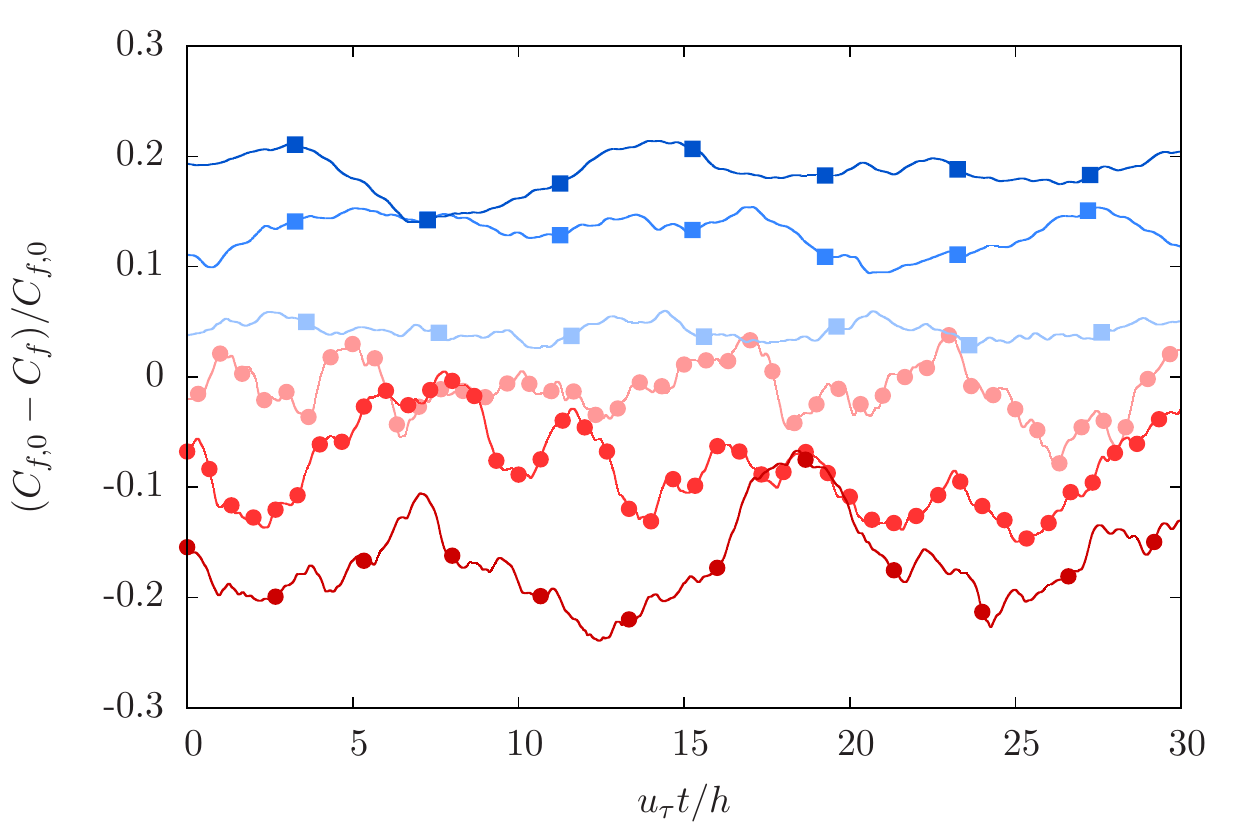}
  \caption{Relative change of skin-friction drag in particle-laden turbulent channel flows. Positive values indicate drag reduction, whereas negative values indicate drag increase. Symbols as in figure \ref{fig:uavg}.\label{fig:friction_coef}
  }
\end{figure}
By modulating the fluid bulk velocity, the dispersed inertial particles lead to a large change in skin-friction drag. Figure \ref{fig:friction_coef} shows times histories of relative change in skin-friction coefficient for all cases in table \ref{tab:parameters}. The reference value, $C_{f,\mathrm{0}}$, corresponds to the skin-friction drag coefficient from the statistically stationary particle-free channel flow. Figure \ref{fig:friction_coef} illustrates how particle inertia plays a selective role by determining the type of flow modulation obtained, be it drag-reducing or drag-increasing, while mass loading acts as an amplifying factor.
For all these cases, we compute the the drag reduction factor  $\mathrm{DR}$, defined as
\begin{equation}
  \mathrm{DR}= \frac{C_{f,\mathrm{0}} - C_f}{C_{f,\mathrm{0}}},
\end{equation}
which takes positive values in the case of drag reduction, and negative values in the case of drag increase. Further, we compute the change in mass flow rate $(\dot{m}_f-\dot{m}_{f,0})/\dot{m}_{f,0}$, where $\dot{m}_{f,0}$ is the mass flow rate in the reference particle-free channel. We report these values in table \ref{tab:drag_reduction}. The greatest drag reduction is obtained with $\Sto^+=30$ particles at the mass loading $M=1.0$, which yields drag reduction factor $\mathrm{DR}=19.54\%$ and corresponding mass flow rate increase of 11.07\%. The latter value is substantially higher than what has been reported in the literature, in particular, by \citet{liNumericalSimulationParticleladen2001} who found an increase in mass flow rate by about $\sim 5\%$ using particles with Stokes number of $O(100)$. This level of drag reduction shows that inertial particles can induce drag reduction at a level comparable with the one obtained using polymer additives \citep{housiadasPolymerinducedDragReduction2003,thaisTemporalLargeEddy2010}, such as in the polymeric channel flow simulations of \citet{housiadasPolymerinducedDragReduction2003} at $\Rey_\tau=180$  where drag reduction $\mathrm{DR}\simeq 25\%$ is reported. Note the drag reduction effect of $\Sto^+=30$ particles decreases significantly at lower mass loadings. At $M=0.2$, these particles reduce drag by only 4.21\%, and yield a modest mass flow rate increase of 2.22\%. This weaker drag reduction is to be expected because the particle feedback force scales with mass loading, thus, particle-induced flow modulation vanishes as $M$ decreases. This also holds for the drag increasing particles with $\Sto^+=6$ whose effect increases with mass loading resulting in drag increase by $16.92\%$ and mass flow rate decrease by 6.10\% at $M=1.0$.

\begin{table}
  \caption{Variation of percent drag reduction and mass flow rate in the present simulations. \label{tab:drag_reduction}}
  \begin{ruledtabular}
    \begin{tabular}{llll}
      Stokes number ($\Sto^+$) & Mass loading ($M$) & DR(\%) & $\Delta \dot{m}_f/\dot{m}_{f,0}$(\%)  \\\hline
      6             & 0.2          & -1.76      & -0.78      \\
                    & 0.6          & -8.90      & -3.93      \\
                    & 1.0          & -16.92      & -6.10      \\
      30            & 0.2          & 4.21       & 2.22      \\
                    & 0.6          & 12.27       & 7.37      \\
                    & 1.0          & 19.54				& 11.07
    \end{tabular}
  \end{ruledtabular}
\end{table}

\subsection{Shear stress balance in the presence of inertial particles}
\label{sec:stress_two_phase}

As with the unladen flow, the structure of the flow in a particle-laden channel results from a balance of stresses applied on the fluid. However, the presence of particles introduces additional stresses that alter the balance in equation (\ref{eq:RSS_1}). To derive a new balance that takes into account particle stresses, we apply Reynolds-averaging to the momentum equation (\ref{eq:NVS_2}). Assuming that particle clustering does not break the dilute limit locally ($1-\phi\simeq 1$), the resulting balance is
\begin{equation}
  \frac{d}{dy}\left(\mu\frac{d }{dy}\langle u_f\rangle - \rho_{f}\langle u_{f}'v_f'\rangle \right)+ \langle F_{p,x}\rangle= -\bigg\langle \frac{\partial p}{\partial x}\bigg\rangle,
    \label{eq:RSS_2}
\end{equation}
where $\langle F_{p,x}\rangle$ represents the mean streamwise particle stresses.
{\color{revision}
The latter can be related to the particle-phase Reynolds shear stress. To do so, we consider the particle conservation equations in the Eulerian frame. Using the Two-Fluid model discussed in \citep{kasbaouiClusteringEulerEuler2019} under the assumption of mono-kinetic particle velocity distribution, the particle mass and momentum conservation equations read
\begin{eqnarray}
    \frac{\partial }{\partial t}(\rho_p\phi) + \nabla \cdot (\rho_p\phi\bm{u}_p)&=& 0 \\
     \frac{\partial}{\partial t}(\rho_p\phi\bm{u}_p) + \nabla \cdot (\rho_p\phi\bm{u}_p\bm{u}_p) &=& -\bm{F}_p +\bm{C}
\end{eqnarray}
where $\bm{u}_p$ is the $\bm{C}$ represents the collision stresses. Neglecting the latter and averaging the streamwise particle momentum balance yields}
\begin{equation}
	\frac{d}{dy}  \left(\rho_p\langle \phi u''_p v''_p\rangle\right)=-\langle F_{p,x}\rangle. \label{eq:RSS_3}
\end{equation}
Here, $u_p''$ and $v_p''$ refer to the streamwise and wall-normal particle-phase velocity fluctuations with respect to the Favre-averaged particle velocities $\widetilde{u}_p=\langle \phi u_p\rangle/\langle \phi\rangle$ and $\widetilde{v}_p=\langle \phi v_p\rangle/\langle \phi\rangle$. Combining equations (\ref{eq:RSS_2}) and (\ref{eq:RSS_3}) yields
\begin{equation}
   \frac{d}{dy}\left(\mu\frac{d}{dy} \langle u_f\rangle - \rho_{f}\langle u_{f}'v_f'\rangle - \rho_{p}\langle \phi u_{p}''v_{p}'' \rangle \right)= -\bigg\langle\frac{\partial p}{\partial x}\bigg\rangle,
    \label{eq:RSS_5}
\end{equation}
which integrates to
\begin{equation}
  \mu\frac{d \langle u_f\rangle }{d y} - \rho_{f}\langle u_{f}'v_f'\rangle - \rho_p\langle \phi u_{p}''v_{p}'' \rangle=\tau_w \left(1-\frac{y}{h}\right) .
    \label{eq:RSS_6}
\end{equation}
Similar to the particle-free channel, equation (\ref{eq:RSS_6}) shows the total stress varies linearly across the channel provided that the particle-phase Reynolds shear stress $\rho_{p}\langle \phi u_{p}''v_{p}'' \rangle $ is also taken into account. Integrating equation (\ref{eq:RSS_6}) twice, leads to an updated expression for the fluid mass flow rate by unit spanwise length which takes into account the effect of the dispersed particles,
\begin{equation}
  \frac{\dot{m_f}}{Lz} = \frac{2}{3}\frac{\tau_w h^2}{\nu}\left(1+\frac{3}{(u_\tau h)^2}\int_{0}^{h}\left(\int_{0}^y \langle u_{f}'v_f'\rangle + \frac{M}{\phi_0} \langle \phi u_{p}''v_{p}'' \rangle dy'\right) dy\right).
    \label{eq:MFR_2}
\end{equation}
The relationship (\ref{eq:MFR_2}) shows that the particles alter the fluid mass flow rate through two competing effects: (i) a direct effect through the particle-phase Reynolds shear stress $\rho_p\langle \phi u_{p}''v_{p}'' \rangle$ which, like the fluid-phase Reynolds shear stress, tends to reduce the mass flow rate, and (ii) an indirect effect through the modulation of the fluid-phase shear stress $\rho_f\langle u_{f}'v_f'\rangle$. It is only when the fluid-phase shear stress is reduced more than can be balanced by the particle-phase Reynolds shear stress that the fluid mass flow rate is increased.

 \begin{figure}\centering
  \centering
  \begin{subfigure}{0.49\textwidth}
    \centering
    \includegraphics[width=3.20in]{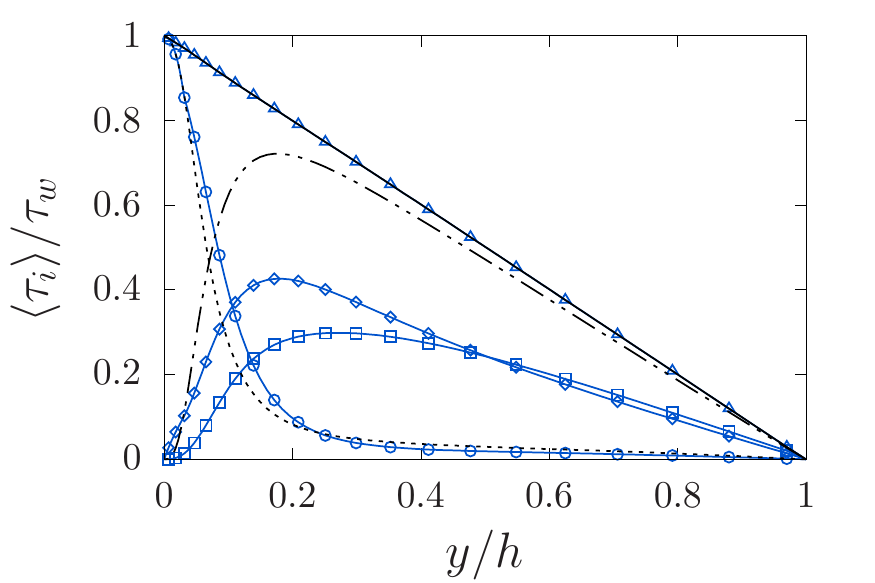}
    \caption{}
    \label{fig:tsscomparest30}
  \end{subfigure}
  \hfill
  \begin{subfigure}{0.49\textwidth}
    \centering
    \includegraphics[width=3.20in]{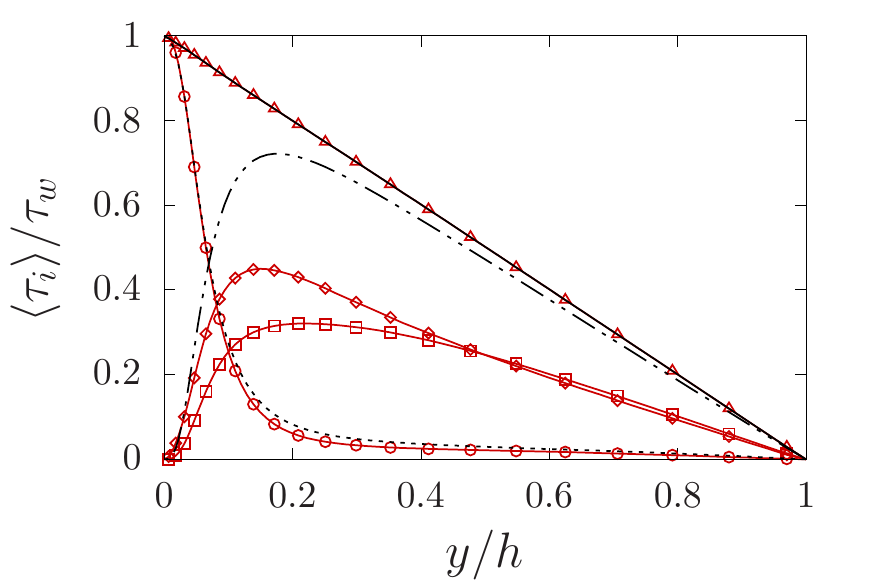}
    \caption{}
    \label{fig:tsscomparest6}
  \end{subfigure}

  \caption{Shear stress contributions as a function of the wall normal distance for the particle-laden turbulent channel flows at (a) $\Sto=30$, $M=1.0$ and (b) $\Sto=6$, $M=1.0$. The viscous shear stress is denoted by (\circles), fluid-phase shear stress by (\squares), particle-phase shear stress by (\diamonds) and the total shear stress by (\triangles). Lines without symbols correspond to the reference single-phase channel as denoted in figure \ref{fig:tss_sp}.}
  \label{fig:tss_comparison}
\end{figure}

Figure \ref{fig:tss_comparison} shows the total shear stress profile and the variations in the fluid and particle stress components for cases $\Sto^+ = 6$ and $\Sto^+ = 30$ at $M = 1.0$. In both cases, the total shear stress varies linearly across the channel as predicted by (\ref{eq:RSS_6}). This first observation validates the two hypotheses underpinning the relationship (\ref{eq:RSS_6}): (i) $1-\phi\simeq 1$ meaning that the particle phase remains dilute even though significant clustering occurs near the walls as we show in \S \ref{sec:mechanism}, and  (ii) collisional stresses are negligible compared to hydrodynamic stresses exerted on particles, even within particle clusters. Considering $\Sto^+ = 30$ particles, figure \ref{fig:tsscomparest30} shows partial relaminarization of the near-wall region. Compared to the reference particle-free flow, the viscous drag increases in the viscous and buffer layers. This modulation is directly linked to the increase in fluid mass flow rate observed in figure \ref{fig:isocontour_u}. Further, the fluid-phase Reynolds shear stresses drops significantly with a peak down to about 39\% of the unladen case, and is shifted further towards the centerline. This drop is partially balanced by  the rise of particle-phase Reynolds shear stress. The latter dominates in the region $0.1\lesssim y/h\lesssim 0.45$ ($18\lesssim y^+\lesssim 81$) and is a comparable to the fluid-phase Reynolds shear stress towards the centerline. Conversely, figure \ref{fig:tsscomparest6} shows that $\Sto^+ = 6$ particles cause a drop of the viscous stress. This is expected since the fluid mass flow rate reduces with these particles. $\Sto^+ = 6$ particles also cause significantly lower fluid-phase Reynolds shear stress, although slightly less than $\Sto^+ = 30$ particles since the peak $\rho_{f}\langle u_{f}'v_f'\rangle$ drops to only 46\% of the unladen case. Further, $\Sto^+ = 6$ particles  cause slightly larger particle-phase shear stress.

\begin{figure}
  \centering
  \begin{subfigure}{0.49\textwidth}
    \centering
    \includegraphics[width=3.20in]{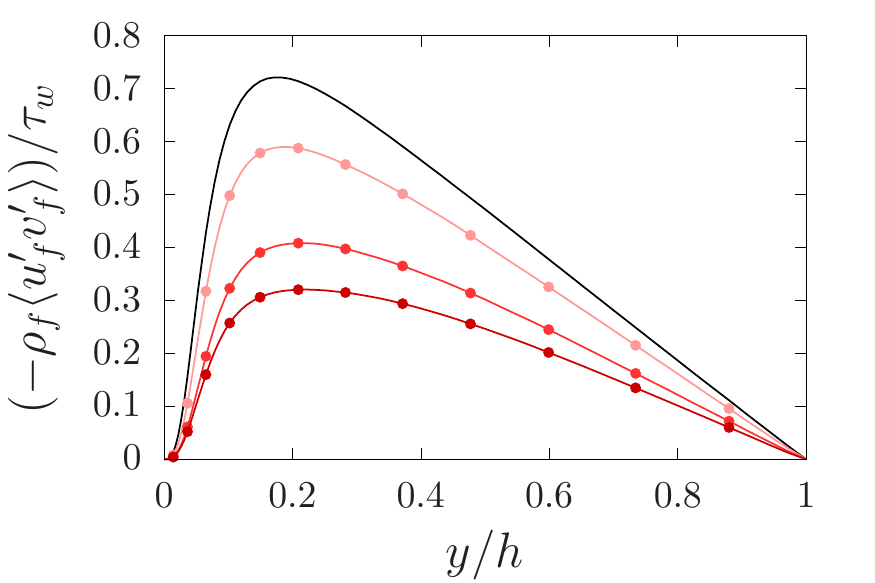}
    \caption{}
    \label{fig:rsscomparest6}
  \end{subfigure}
  \hfill
  \begin{subfigure}{0.49\textwidth}
    \centering
    \includegraphics[width=3.20in]{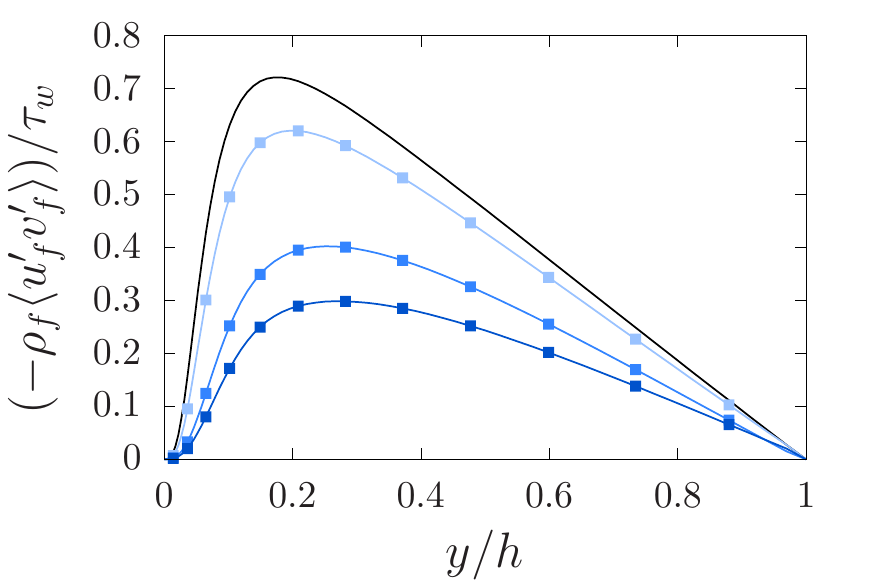}
    \caption{}
    \label{fig:rsscomparest30}
  \end{subfigure}\\
	\begin{subfigure}{0.49\textwidth}
		\centering
		\includegraphics[width=3.20in]{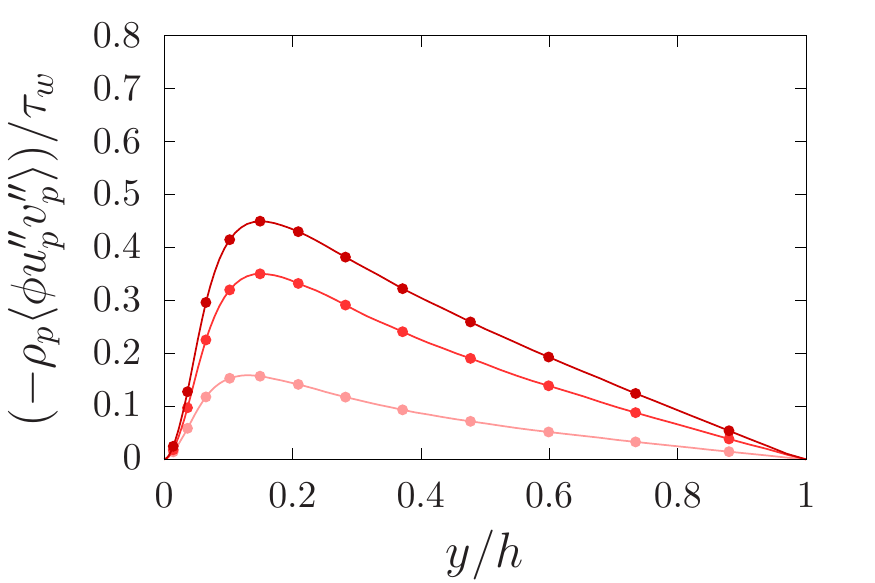}
		\caption{}
		\label{fig:psscomparest6}
	\end{subfigure}
	\hfill
	\begin{subfigure}{0.49\textwidth}
		\centering
		\includegraphics[width=3.20in]{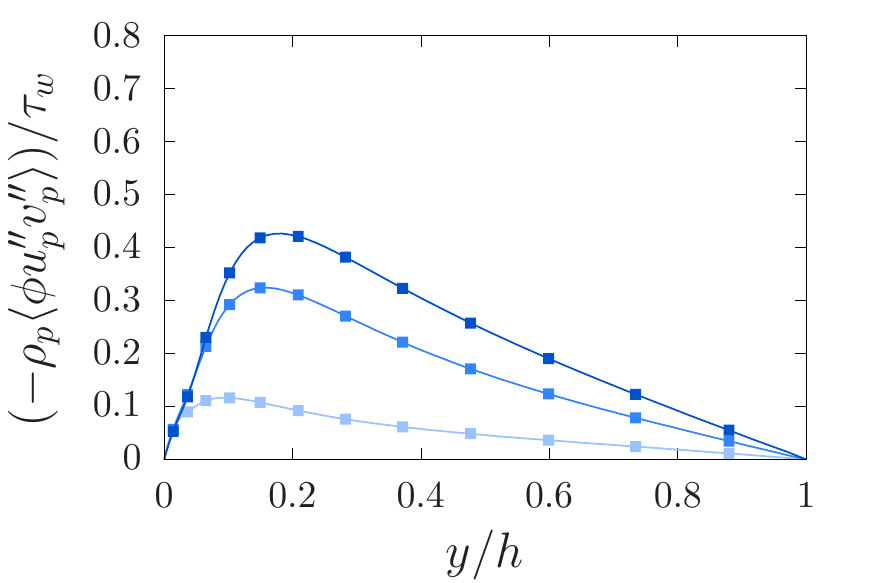}
		\caption{}
		\label{fig:psscomparest30}
	\end{subfigure}
  \caption{Fluid-phase and particle-phase Reynolds shear stress for (a,c) $\Sto^+=6$ and (b,d) $\Sto^+=30$ respectively. Symbols as in figure \ref{fig:uavg}. The solid black line denotes the single-phase Reynolds shear stress.}
  \label{fig:rss_pss_comparison}
\end{figure}

\begin{figure}
	\centering
	\begin{subfigure}{0.49\textwidth}
		\centering
		\includegraphics[width=3.20in]{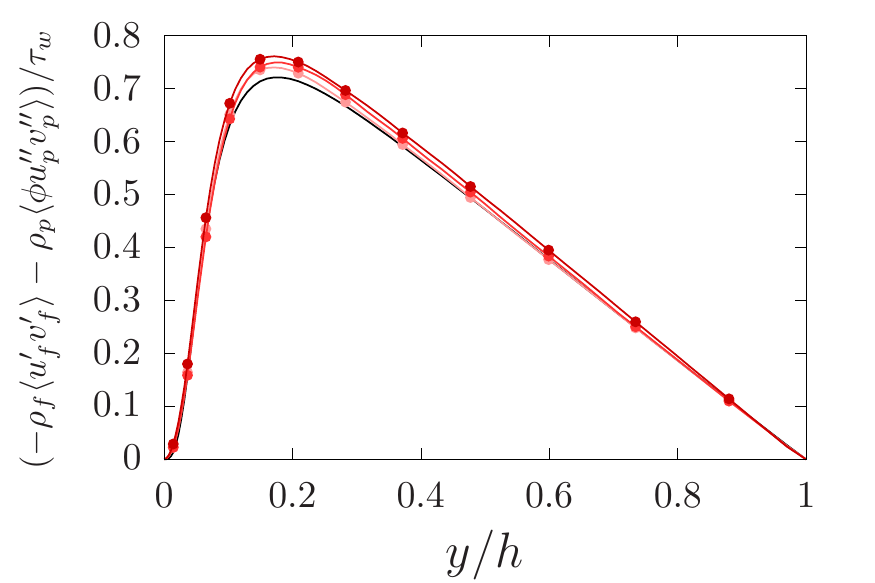}
		\caption{}
		\label{fig:sscomparest6}
		\end{subfigure}\hfill
		\begin{subfigure}{0.49\textwidth}
		\centering
		\includegraphics[width=3.20in]{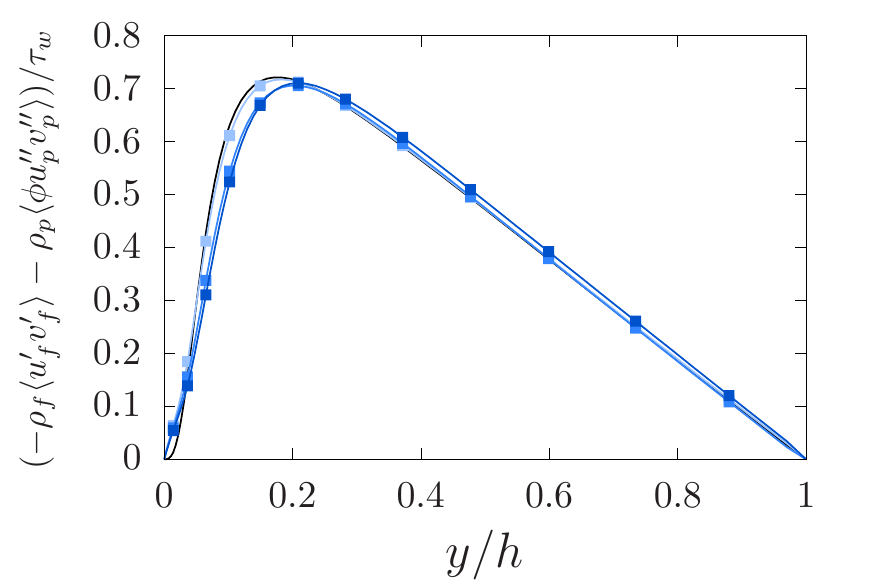}
		\caption{}
		\label{fig:sscomparest30}
\end{subfigure}
\caption{Total Reynolds shear stress for (a) $\Sto^+=6$ and (b) $\Sto^+=30$ respectively. Symbols as in figure \ref{fig:uavg}. The solid black line denotes the single-phase Reynolds shear stress.}
\label{fig:rss_pss_sum}
\end{figure}

Figure \ref{fig:rss_pss_comparison} shows the effect of varying mass loading on the fluid and particle shear stresses. For both $\Sto^+=30$ and $\Sto^+=6$ particles, the particle shear stress rises with increasing mass loading while the fluid Reynolds shear stress drops. As shown by the relationships (\ref{eq:RSS_6}) and (\ref{eq:MFR_2}), the competition between increasing particle shear stress and reducing fluid Reynolds shear stress is what ultimately determines whether the particles increase or decrease the mass flow rate, and \emph{a fortiori}, drag reduction or drag increase, respectively. Figure \ref{fig:rss_pss_sum} shows how increasing mass loading causes a progressive deviation of the total Reynolds shear stress  $\rho_{f}\langle u_{f}'v_f'\rangle + \rho_p \langle \phi u_{p}''v_{p}'' \rangle$ from the single phase Reynolds shear stress. It is clear that $\Sto^+=30$ particles reduce the total Reynolds shear stress, although at a rate that varies little from $M=0.6$ to $M=1.0$ suggesting a possible saturation. With $\Sto^+=6$ particles, there is an increase of total Reynolds shear stress which accentuates with increasing mass loading.

\subsection{Interplay between particle clusters and near-wall coherent structures}
\label{sec:mechanism}
In this section, we show that modulating the skin-friction drag depends to a large extent on how particle clusters interact with near-wall coherent structures.

\begin{figure}
  \includegraphics[width=5in]{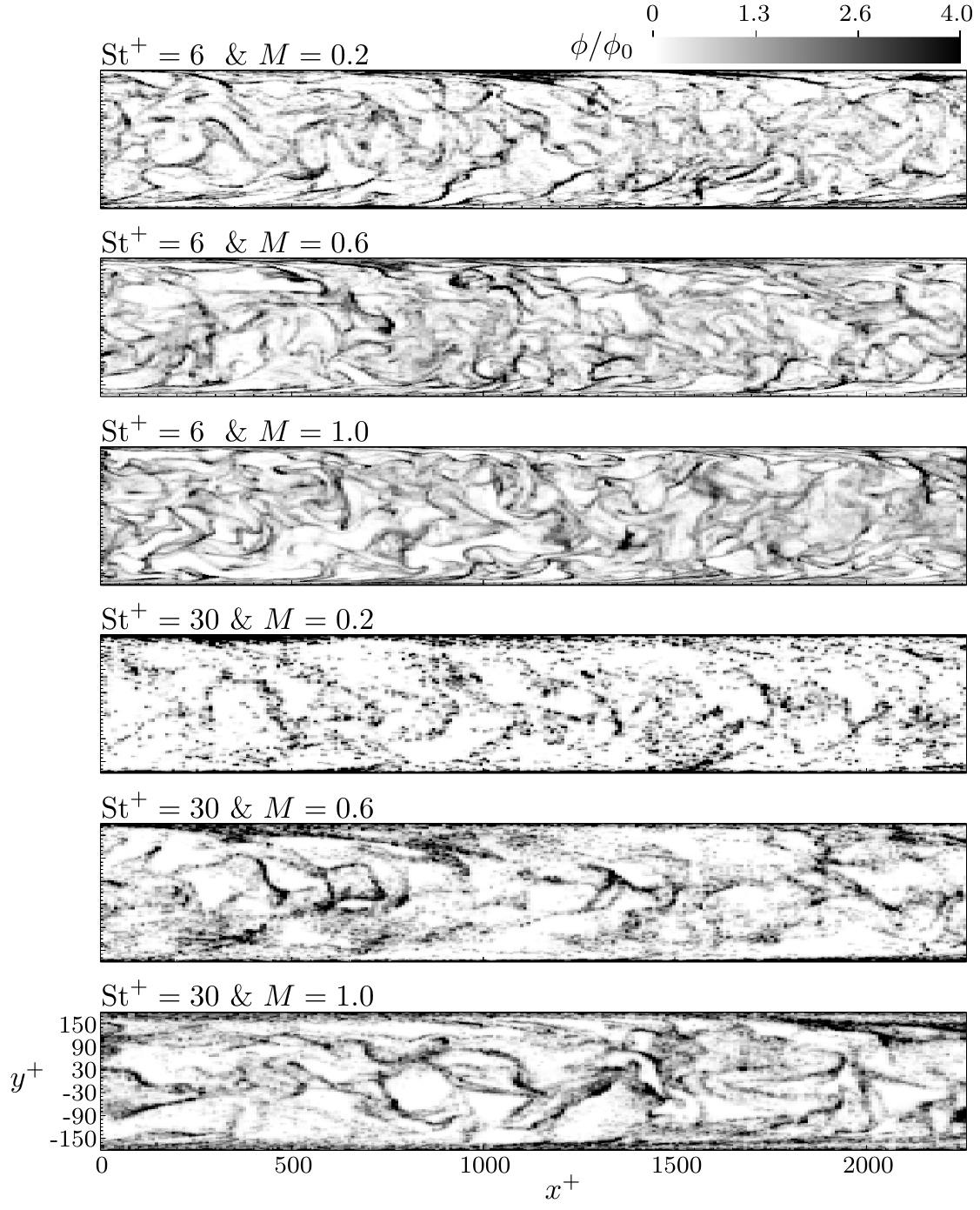}
  \caption{Isocontours of normalized particle volume fraction in a wall-normal plane showing the presence of clusters and the accumulation of particles near the walls. As in figure \ref{fig:isocontour_u}, the larger domain for $\Sto^+ = 30$ particles is truncated to the same size as the domain for $\Sto^+ = 6$ particles to facilitate visual comparison.}
  \label{fig:vfp}
\end{figure}
The distribution of $\Sto^+=6$ and $\Sto^+=30$ particles within the channel is strongly inhomogeneous.
Visualization of normalized particle volume fraction in a wall-normal plane in figure \ref{fig:vfp} shows that the particles concentrate in long filamentous clusters that may span the entire channel height.
$\Sto^+=30$ particles form clusters that are relatively denser and further elongated in the streamwise direction compared to clusters formed by $\Sto^+=6$ particles. Figure \ref{fig:vfp} also shows that the normalized particle volume fraction within the bulk of the channel is lower at mass loading $M=0.1$, compared to the bulk normalized volume fraction at $M=0.6$ and 1.0. This points to a tendency of particles to accumulate near the walls that is stronger at $M=0.1$  than at $M=0.6$ and $M=1$.  Note that the formation of such clusters is expected owing to the fact that the particles considered in this study have significant inertia. As previously discussed by several investigators, inertial particles in wall-bounded turbulent flows tend to form clusters due to two effects, namely, turbophoresis, i.e., the migration of inertial particles to lower turbulence regions near the walls \citep{caporaloniTransferParticlesNonisotropic1975,reeksTransportDiscreteParticles1983,nowbaharTurbophoresisAttenuationTurbulent2013,kuertenTurbulenceModificationHeat2011}, and preferential concentration, i.e., the migration of inertial particles from vortical regions to straining regions of the flow \citep{eatonPreferentialConcentrationParticles1994,marchioliStatisticsParticleDispersion2008,kasbaouiTurbulenceModulationSettling2019,fongVelocitySpatialDistribution2019}. It follows that the particle feedback force is concentrated along these structures, and that the resulting flow modulation depends largely on the cluster morphology and dynamics.

\begin{figure}\centering
  \begin{subfigure}{0.49\linewidth}
    \centering
    \includegraphics[width=3.2in]{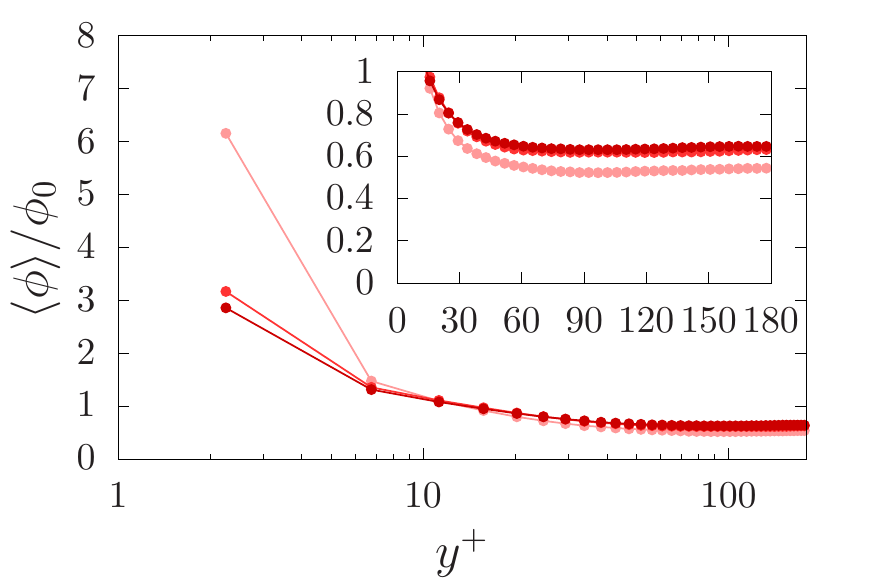}
    \caption{}
    \label{fig:pnd_st6}
  \end{subfigure}\hfill
  \begin{subfigure}{0.49\linewidth}
    \centering
    \includegraphics[width=3.2in]{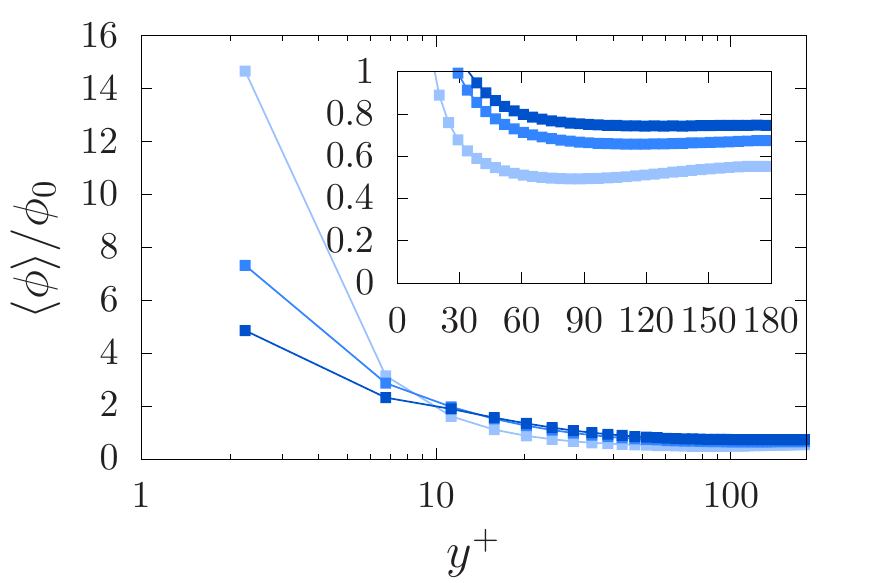}
    \caption{}
    \label{fig:pnd_st30}
	\end{subfigure}
  \caption{Particle number density normalized by the average particle number density as a function of the wall normal distance for (a) $\Sto^+ = 6$ and (b) $\Sto^+ = 30$ at various mass loadings. Symbols as in figure \ref{fig:uavg}.}
  \label{fig:PND}
\end{figure}

Although particle clusters can be observed throughout the channel, it is near the walls that the majority of particles accumulate. Figure \ref{fig:PND} shows the variation of the normalized plane-averaged volume fraction ${\langle \phi\rangle}/ \phi_0$ with the wall normal distance. Within the region $y^+ < 10$, the local particle volume fraction is several times larger than the mean volume fraction $\phi_0$, which shows that the majority of the particles accumulate near the walls. $\Sto^+=30$ particles lead to the largest wall accumulation reaching ${\langle \phi\rangle}/ \phi_0\simeq 4.98$ at $M=1$ compared to ${\langle \phi\rangle}/ \phi_0\simeq 2.62$ for $\Sto^+=6$ particles at the same mass loading. Similar observations were made by \citet{nilsenVoronoiAnalysisPreferential2013} and \citet{yuanThreedimensionalVoronoiAnalysis2018} who, despite considering only one-way coupling, found that particles with $\Sto^+=30$ have the greatest wall-accumulation among particles with $\Sto^+$ in the range 1-100. Interestingly, the particle wall accumulation reduces when mass loading increases. At $M=0.1$, the particle volume fraction at the wall rises to $\langle\phi\rangle/\phi_0\simeq 14.96$ and 6.7 for $\Sto^+=30$ and $\Sto^+=6$, respectively. This finding is in agreement with the observation from figure \ref{fig:vfp} that the relative bulk particle volume fraction is lowest at $M=0.1$ as relatively more particles accumulate at the walls with decreasing $M$. This effect likely results from two-way coupling, given that particle-particle collisions are weak in the present semi-dilute regime.

Here, we stress that capturing the particle ropes accurately and the subsequent flow modulation requires much larger domains than those generally used in simulations of particle-laden turbulent channel flows \citep{rousonPreferentialConcentrationSolid2001,zhaoTurbulenceModulationDrag2010a,bernardiniReynoldsNumberScaling2014,costaInterfaceresolvedSimulationsSmall2020,jieExistenceFormationMultiscale2022}. The present large domain used for simulations with $\Sto^+=30$ particles is sufficiently wide to allow a natural development of flow and particle structures in the spanwise direction. However, even with a streamwise length of $12\pi h\sim 38h$, the domain remains too short to properly characterize the average streamwise length of the particle ropes.

\begin{figure}
    \centering
  \begin{subfigure}{\linewidth}
    \centering
    \includegraphics[width=4.5in]{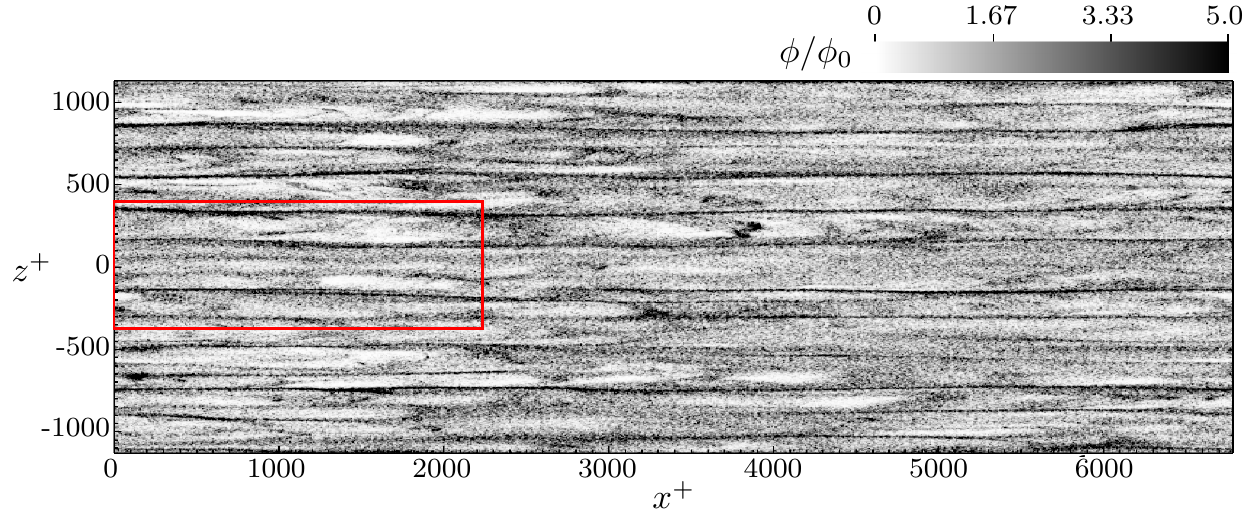}
    \caption{}
    \label{fig:cluster_length_st30_2d1}
  \end{subfigure}
  \begin{subfigure}{\linewidth}
    \centering
    \includegraphics[width=4in]{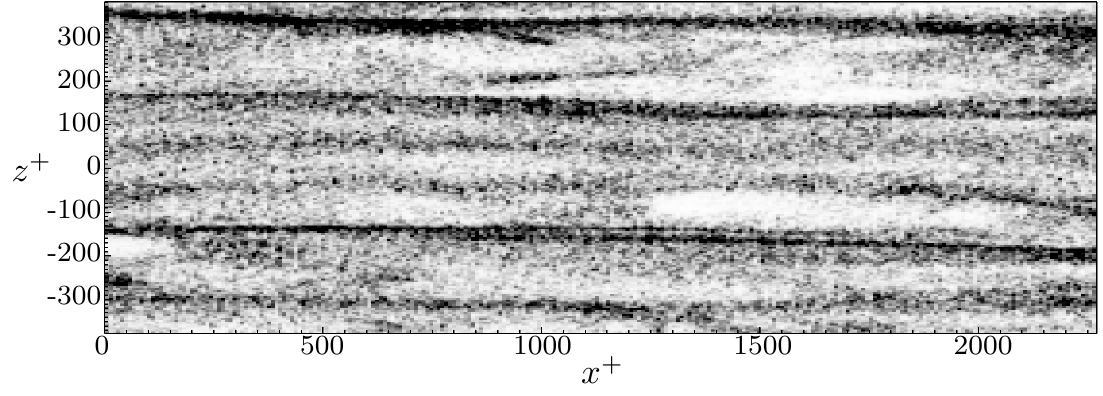}
    \caption{}
    \label{fig:cluster_length_st30_2d2}
  \end{subfigure}
  \begin{subfigure}{\linewidth}
    \centering
    \includegraphics[width=4in]{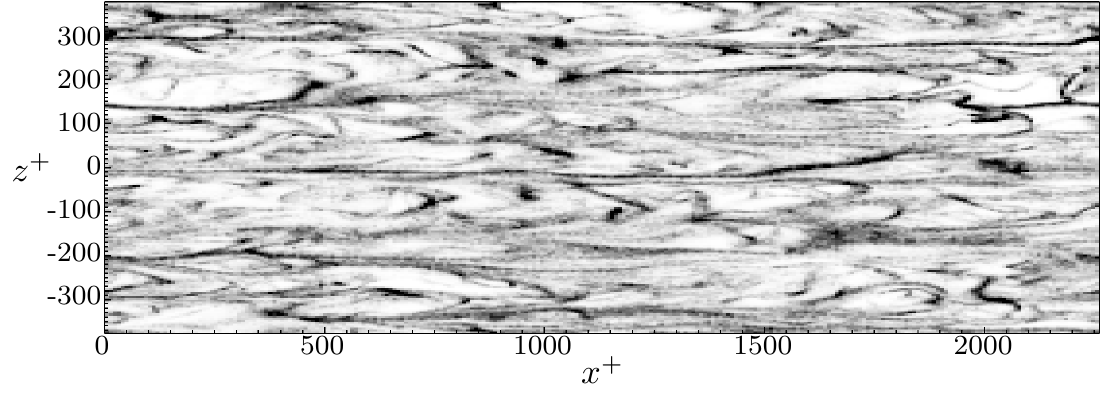}
    \caption{}
    \label{fig:cluster_length_st6_2d1}
\end{subfigure}
  \caption{\textcolor{revision}{Isocontours of normalized particle volume fraction at $y^+ = 10$ for (a,b) $\Sto^+ = 30, M = 1.0$ and (c) $\Sto^+ = 6, M = 1.0$. The view in (b) corresponds to the area marked by the red rectangle in (a).}}
  \label{fig:cluster_length}
\end{figure}
With most of the particles concentrating near the walls, clusters found therein have the largest impact on the carrier flow. As shown in figure \ref{fig:cluster_length}, the topology of these structures varies significantly depending on whether the particles are drag-reducing ($\Sto^+=30$) or drag-increasing ($\Sto^+=6$). For better comparison of the scales, figure \ref{fig:cluster_length_st30_2d2} shows a view of the particle volume fraction field for $\Sto^+=30$ particles cropped to the same dimensions as the smaller domain used with $\Sto^+=6$ particles and shown in figure \ref{fig:cluster_length_st6_2d1}. In contrast with $\Sto^+=6$ particles, the higher inertia particles at $\Sto^+=30$ form distinctively long and stable clusters. These structures, which we call \emph{ropes}, span the entire length of the domain in the streamwise direction, i.e, over 6000 wall units. The ropes travel downstream but remain stable and coherent for dynamically significant times. Further, the ropes repeat periodically in the spanwise direction in a fashion reminiscent of low-speed streaks discussed in \S\ref{sec:unladen}. This suggests that formation of these ropes results from the interaction of particle clusters with coherent flow structures in the buffer layer. The fact that no such ropes are observed with $\Sto^+=6$ particles suggests that intermittent flow structures in the buffer layer are capable of breaking down clusters formed by low inertia particles, whereas clusters formed by particles with large inertia retain their spatial and temporal coherence. The stable particle ropes may in turn alter the near-wall coherent flow structures.

\begin{figure}
    \centering
    \begin{subfigure}{\linewidth}
      \centering
      \includegraphics[width=4.5in]{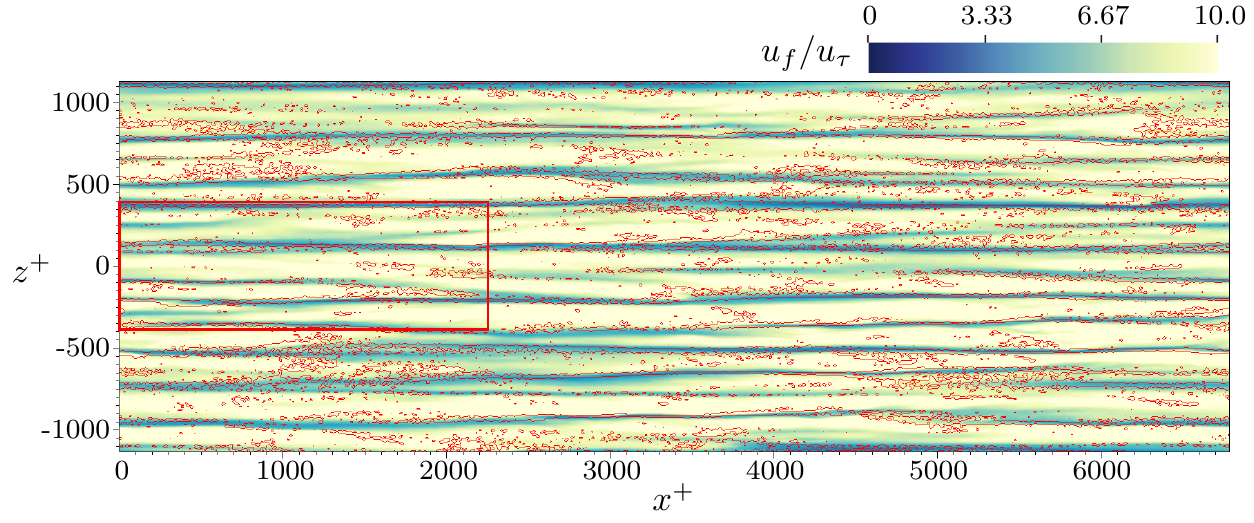}
      \caption{$\Sto^+ =30, M = 1.0$}
      \label{fig:riblet_S30M10_large}
    \end{subfigure}
    \begin{subfigure}{\linewidth}
      \centering
      \includegraphics[width=4in]{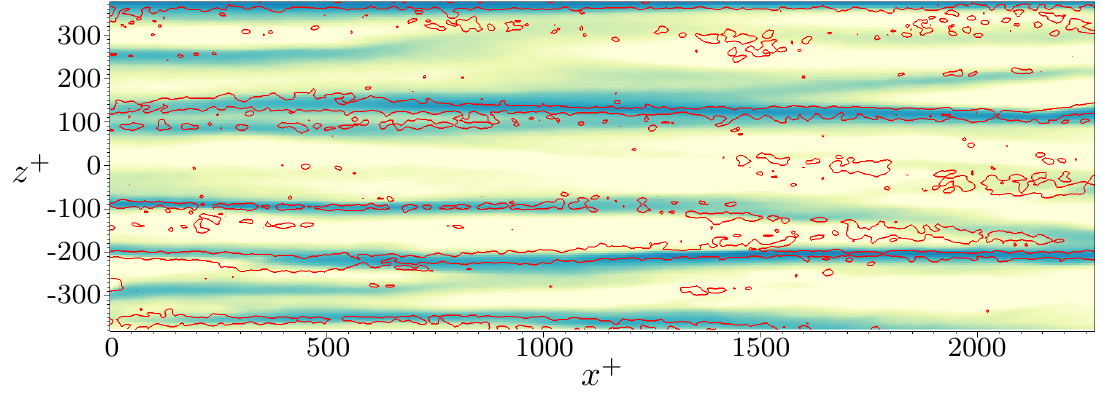}
      \caption{$\Sto^+ =30, M = 1.0$}
      \label{fig:riblet_S30M10}
    \end{subfigure}
    \begin{subfigure}{\linewidth}
    \centering
    \includegraphics[width=4in]{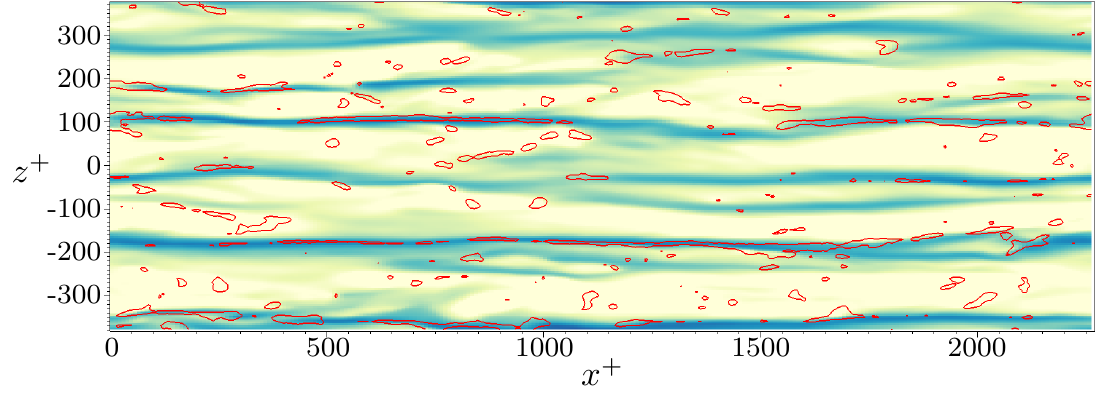}
    \caption{$\Sto^+ = 6, M = 1.0$}
    \label{fig:riblet_S6M10}
  \end{subfigure}
  \caption{\textcolor{revision}{Overlay of the isocontours of fluid streamwise velocity, and the contour of the relative particle volume fraction $\phi/\phi_0 = 3$ at $y^+ = 10$ for (a,b) $\Sto=30$, $M=1.0$ and (c) $\Sto^+ = 6, M = 1.0$. The view in (b) corresponds to the area marked by the red rectangle in (a).}}
  \label{fig:riblet_comparison}
\end{figure}
In order to shed light on how particle ropes interact with near-wall coherent flow structures, we report in figure  \ref{fig:riblet_comparison} isocontours of streamwise velocity at $y^+ = 10$ with the iso-level $\phi=3\times\phi_0$ overlayed on top. The latter shows the regions where the particles cluster. For the flow laden with $\Sto^+ = 30$ particles at $M=1$, we observe that the long ropes align well with the low-speed streaks, showing that the dynamics of these two coherent structures are interlinked. Compared to the unladen flow (see figure \ref{fig:streak_SP_a}), the low-speed streaks are visibly further elongated in a way similar to how the particle ropes extend in the streamwise direction. The spanwise spacing of the low-speed streaks also increases and appears comparable to the spanwise spacing of the ropes. In the case of the flow laden with $\Sto^+=6$ particles at $M=1$, the clusters are also primarily found in the low-speed streaks. However, the streamwise length of these clusters is much shorter in comparison with the low-speed streaks and with the ropes formed by $\Sto^+ = 30$ particles. In addition, the streamwise length and spanwise spacing of low-speed streaks increase compared to the particle-free flow, although not to the same extent as with $\Sto^+=30$ particles.

\begin{figure}
	\centering
	\begin{subfigure}{0.45\linewidth}
		\includegraphics[width=\linewidth]{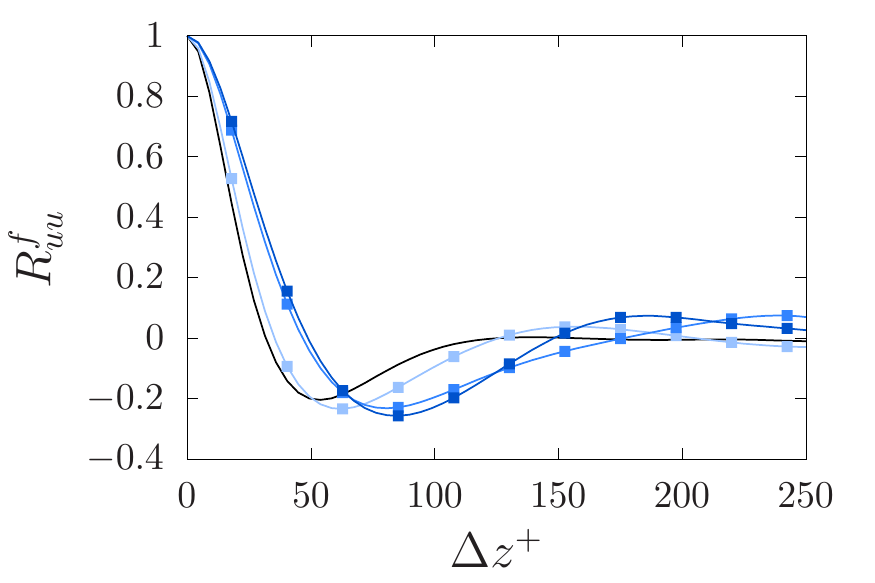}
		\caption{\label{fig:spacing_st30_f}}
	\end{subfigure}
	\hfill
	\begin{subfigure}{0.45\linewidth}
		\includegraphics[width=\linewidth]{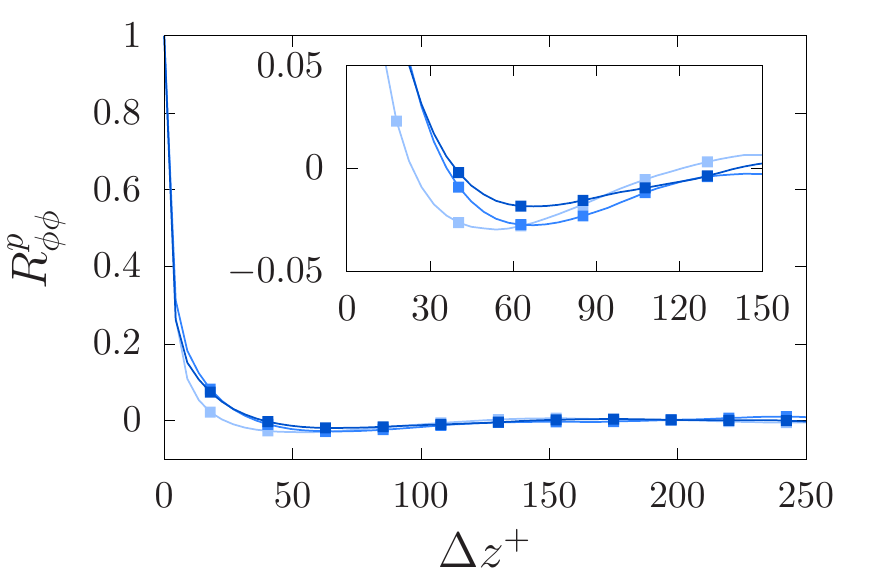}
		\caption{\label{fig:spacing_st30_p}}
	\end{subfigure}
	\begin{subfigure}{0.45\linewidth}
		\includegraphics[width=\linewidth]{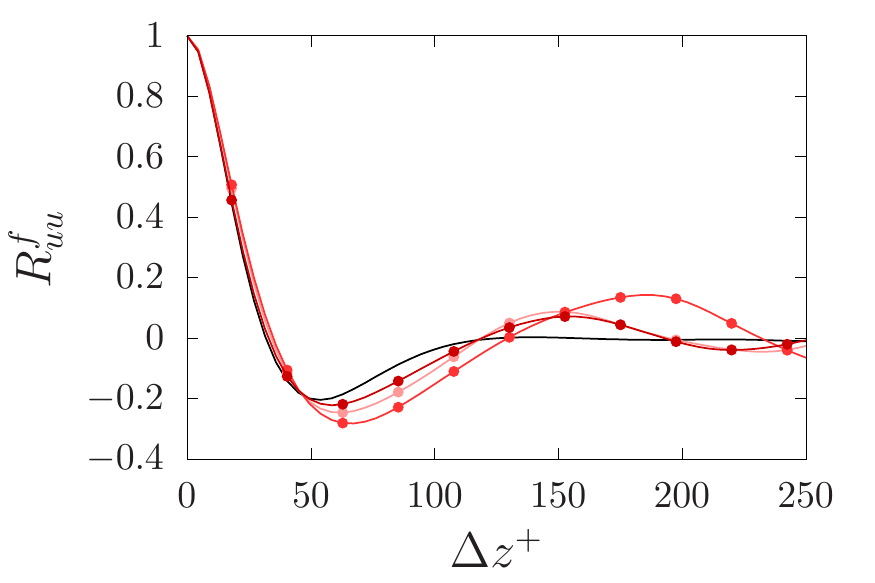}
		\caption{\label{fig:spacing_st30_p}}
	\end{subfigure}
	\hfill
	\begin{subfigure}{0.45\linewidth}
		\includegraphics[width=\linewidth]{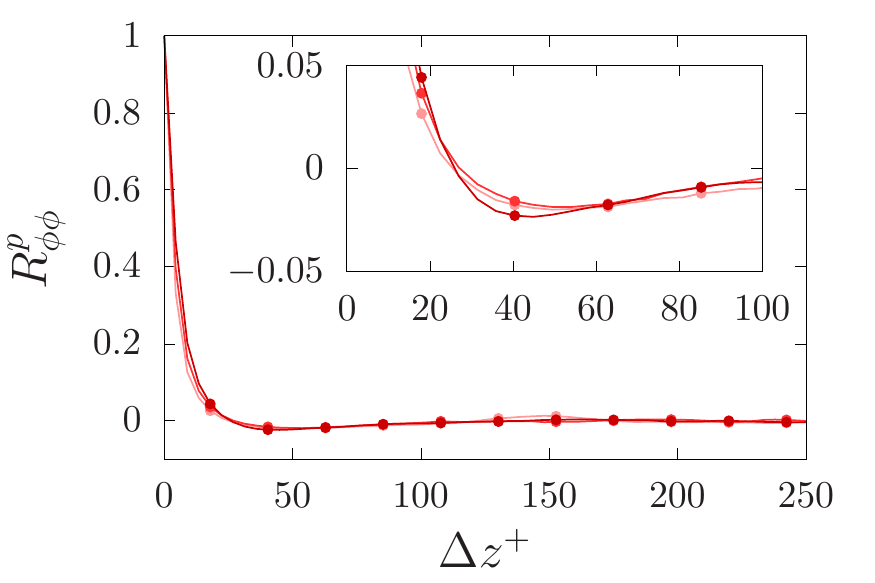}
		\caption{\label{fig:spacing_st6_p}}
	\end{subfigure}
	\caption{Variation with spanwise spacing of the two-point autocorrelation of the streamwise fluid fluctuations and particle volume fraction fluctuations in the spanwise direction for the (a,b) drag-reducing case $\Sto^+=30$ (\textcolor{blue}{\protect\scalebox{1.25}{$\blacksquare$}}) and (c,d) drag increasing case $\Sto^+ = 6$ (\textcolor{red}{\protect\scalebox{1.75}{$\bullet$}}). Darker symbols correspond to larger mass loading which varies from 0.2 to 1.0. The solid black line represents the particle-free channel flow.
	\label{fig:spacing}}
\end{figure}
To characterize quantitatively the spanwise spacing of particle clusters and their impact on the spanwise spacing of low-speed streaks, we compute the two-point autocorrelation of the particle volume fraction fluctuations,
\begin{equation}
	    R^p_{\phi\phi}(\Delta z;y_0)=\frac{\langle{\phi' (x,y_0,z,t)\phi' (x,y_0,z+\Delta z,t)}\rangle}{\langle{\phi'^2}\rangle},
\end{equation}
and the the two-point autocorrelation of the streamwise velocity fluctuations $R^f_{uu}$. Figure \ref{fig:spacing} shows the variation $ R^p_{\phi\phi}$ and $ R^f_{\phi\phi}$ with spanwise spacing at $y^+=10$. Similar to how the low-speed streak spacing $\lambda_f^+$ is defined, we define $\lambda_p^+$, the spanwise spacing of particle clusters, as twice the distance between the origin and $\Delta z^+$ where $R^p_{\phi\phi}$ reaches a first minimum.

\begin{table}
  \caption{Spanwise spacing of the low-speed streaks and particle ropes. \label{tab:spacing}}
  \begin{ruledtabular}
    \begin{tabular}{llll}
      Stokes number ($\Sto^+$) & Mass loading ($M$) & $\lambda^+_f$ & $\lambda^+_p$ \\\hline
      (Particle-free) & 0 & 106 & $-$ \\
      6             & 0.2          & 125      & 99     \\
                    & 0.6          & 134      & 108    \\
                    & 1.0          & 116      & 90     \\
      30            & 0.2          & 126      & 108    \\
                    & 0.6          & 161      & 130    \\
                    & 1.0          & 170	  & 135    
    \end{tabular}
  \end{ruledtabular}
\end{table}

Table \ref{tab:spacing} shows the values of $\lambda^+_f$ and $\lambda^+_p$ for all cases simulated.
For the drag-reducing cases at $\Sto^+ = 30$, it is clear that as the mass loading is increased from 0.2 to 1.0 the low-speed streak spanwise spacing increases from $\lambda^+_f = 126$ to $170$. These are significant increases compared to the low-speed streak spacing of $\lambda^+_f = 106$ in the particle-free channel. The rope spacing $\lambda_p^+$ increases from $\lambda^+_p = 108$ to $135$ as mass loading is increased. 
\textcolor{revision}{The disparity between $\lambda_p^+$ and $\lambda_f^+$ is likely due to small particle clusters that detach from the main ropes due to the spanwise meandering of ropes and low-speed streaks.}
%
In comparison, $\Sto^+ = 6$ particles lead to substantially lower modulation of the low-speed streaks. As shown in table \ref{tab:spacing}, the spanwise spacing of the low-speed streaks varies between $\lambda^+_f = 116$ and $134$ when $\Sto^+ = 6$ particles are dispersed. The corresponding spacing of particle clusters varies in the range of $\lambda^+_p = 90-108$, \textcolor{revision} {with less disparity between $\lambda^+_p$ and $\lambda^+_f$ compared to the flow laden with $\Sto^+ = 30$ particles. This suggests that $\Sto^+ = 6$ clusters are more closely aligned with the high-strain low-vorticity regions found within the low-speed streaks, likely due to their lower inertia.}
%

Note that the two-way coupling plays a critical role in the arrangement of low-speed streaks and particle clusters. In a prior study by \citet{jieExistenceFormationMultiscale2022}, where the authors considered one-way coupled Euler-Lagrange simulations of particle-laden channel flows at $\Rey_\tau$ between 600 and 2000, the absence of feedback force from the particles leads to low-speed streaks that have identical characteristics to those of a particle-free turbulent channel flow. The data presented by the authors further suggests that the particle cluster spanwise spacing varies little with Reynolds number and is about $\lambda_p^+ \sim 114$ for $\Sto^+ = 30$ particles. However, as we have shown in this study,  $\lambda_p^+$ and $\lambda_f^+$ reach considerably higher values when two-way coupling is significant since the dynamics of clusters and near-wall coherent structures become more inter-dependent.

\begin{figure}
  \centering
  \includegraphics[width=4in]{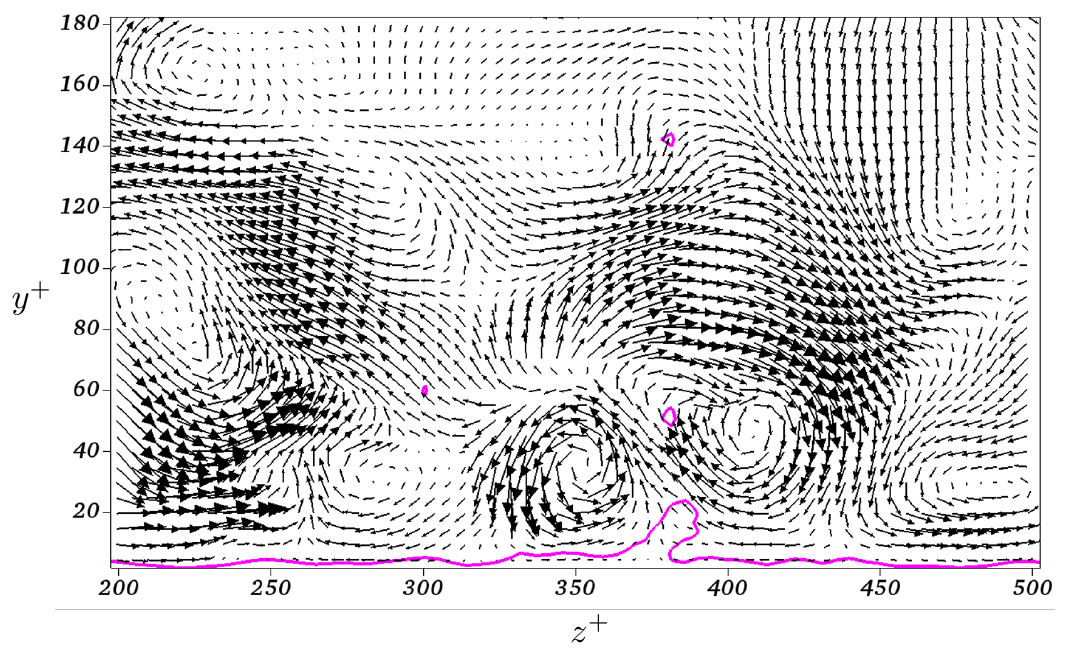}
  \caption{\textcolor{revision}{Instantaneous velocity vectors overlayed by contour of particle volume fraction $\phi/\phi_0 = 3$, for the case $\Sto^+ = 30, M = 1.0$, show particle ropes forming in the high strain region between quasi-streamwise vortices.}}
  \label{fig:vorticity}
\end{figure}

Figure \ref{fig:vorticity} shows an example of how $\Sto^+=30$ particles are distributed in the vicinity of a pair of quasi-streamwise vortices. The particles form ropes by concentrating in the straining region between the pair of vortices, consistently with the preferential concentration mechanism. Pockets of particles can be seen ejected upward towards the centerline, which results in a downward feedback force on the fluid. This process is self-sustaining because the ejected particles eventually return to the near-wall region due to turbophoresis, where they accumulate again along particle ropes. The feedback force from these clusters contributes to the the suppression of bursting and stabilization of quasi-streamise vortices. Consequently, low-speed streaks nested in-between quasi-streamwise vortices extend further than possible in particle-free flows. Because bursting events contribute largely to the Reynolds shear stress production \citep{willmarthStructureReynoldsStress1972}, the stabilizing role of  $\Sto^+=30$ particles is likely the main reason these particles reduce the fluid-phase Reynolds shear stress to the extent shown in \S \ref{sec:stress_two_phase}, and \emph{in fine}, skin-friction drag reduction.

\section{Discussion and conclusion}
\label{sec:conclusion}
We have shown that it is possible to induce significant drag reduction using mono-dispersed spherical particles, provided that their inertia and concentration are tuned appropriately. Using four-way coupled Euler-Lagrange simulations of semi-dilute particle-laden turbulent channel flows at $\Rey_\tau=180$ with mass loading varying between $M=0.2$ and 1.0, we have shown that the particle Stokes number is a determining factor in the type of modulation observed. Among the two types of particles we have considered having friction Stokes number $\Sto^+=6$ or 30, drag increase is observed with the former, and drag reduction is observed with the latter. Mass loading plays an amplifying role in such a way that at $M=0.2$ the drag increase or decrease observed is negligible, whereas these effects become significant at $M=1.0$, resulting in drag reduction factors of up to $\mathrm{DR}=19.74\%$ and $\mathrm{DR}=-16.92\%$ for $\Sto^+ = 30$ and $\Sto^+ = 6$, respectively. A key observation is that particle clusters and coherent structures found in the near-wall region have tightly coupled evolutions. Modifications to the latter by the particle clusters explain in part the observed changes to skin-friction drag.

For the drag-reducing cases considered, the largest drag reduction is achieved for the case $\Sto^+ = 30, M = 1.0$ in which skin friction drag drops by $\mathrm{DR} = 19.54\%$ and mass flow rate increase by $\Delta \dot{m}_f/\dot{m}_{f,0} = 11.07\%$ compared to the reference particle-free flow. A distinct feature visually observed for particles at $\Sto^+ = 30$ is the existence of concentrated clusters along the channel walls with local particle volume fraction several times larger than the mean. These clusters, that we call \emph{ropes}, are very long structures that span the entire domain in the streamwise direction, about $38h$. Further, the ropes appear to preferably align with the low-speed streaks of the flow, and to cause their structure to differ considerably from those found in particle-free flows. The observed modulation which includes a stabilization of the low-speed streaks, reduction in bursting, elongation in the streamwise direction, and increase in spanwise separation result from the collective feedback force from particles located within these concentrated ropes. Using two-point autocorrelations, we found that the low-speed streaks spanwise spacing $\lambda_f^+$ increases from the little varying value $\lambda_f^+=106$ in particle-free flows to $\lambda_f^+=170$ when the flow is laden with $\Sto^+ = 30$ particles at $M = 1.0$. In comparison, the ropes spacing in this case is $\lambda_p^+ = 135$. The disparity between $\lambda_f^+$ and $\lambda_p^+$ 
\textcolor{revision}{is likely due to small clusters that detach from the main ropes due to the spanwise meandering of ropes and low-speed streaks.}
%
%
While dispersed particles cause additional stresses on the fluid, the modulation of near-wall coherent structures by $\Sto^+ = 30$ particles leads to greater reduction in Reynolds shear stress, which ultimately causes a partial relaminarization of the near-wall flow and skin-friction drag reduction.

In contrast to the larger inertia particles, dispersing $\Sto^+ = 6$ particles in the flow causes drag increase. The largest effect is observed at $M = 1.0$ which yields a drag increase of $16.92\%$ and mass flow rate decrease of $6.10\%$. These lower inertia particles do not show the same type of clustering seen with $\Sto^+ = 30$ particles. Particle cluster sizes are smaller and no rope-like clusters spanning the entire length in the streamwise direction are observed. Furthermore, the low-speed streaks also do not seem to widen or elongate at the rate that was observed for the $\Sto^+ = 30$ case. While the low-speed streak spacing increases compared to the particle-free case, from $\lambda^+_f = 106$ to $116$ at the highest drag increasing case at $\Sto^+ = 6$ and $M = 1.0$, the change is significantly lower when compared to the $\Sto^+ = 30$ case. The spanwise spacing of the particle clusters is also significantly lower with $\lambda^+_p = 90-108$. For these $\Sto^+ = 6$ particles, the low-speed streaks and particle clusters are more closely aligned. This is because lower inertial particles are less likely to escape the low-speed regions where they are mostly located. While they do exert a feedback force that reduces near-wall coherent structures, the resulting drop in Reynolds shear stress is not sufficient to balance the additional stresses exerted by the particles, hence leading to drag increase.

We shall note that the mechanisms discussed in this study hold some similarities with those found in polymeric flows. Here, friction Stokes number, ratio of the particle response time and friction time scale, is analogous to the Weissenberg number,  ratio of the polymer elasticity timescale and the friction time scale. Like the Weissenberg number in polymeric flows, the Stokes number determines whether drag reduction or drag increase is achieved. Stresses induced by inertial particles are analogous to stresses resulting from polymers. In both cases, drag reduction is determined by the extent to which the fluid Reynolds shear stress is suppressed in comparison to the additional particle or polymer stresses. However, polymers modulate flow structures through contraction and elongation, whereas inertial particles act on the flow through their drag force. Further, the mechanisms related to particle clustering, rope formation, and interplay with near-wall coherent structures are unique to particle-laden flows.

Finally, the fact that $\Sto^+ = 6$ and  $\Sto^+ = 30$ particles lead to opposite drag modulation suggests that there is a critical Stokes number above which drag reduction is obtained. This threshold may depend on mass loading and density ratio. Moreover, while we have shown close to 20\% drag reduction using $\Sto^+ = 30$ particles at mass loading $M=1$, varying Stokes number may lead to higher drag reduction. Additional simulations are required to find the threshold Stokes number for drag reduction and establish a regime map of drag modulation. 

\section*{Acknowledgement}
Acknowledgement is made to the donors of the American Chemical Society Petroleum Research Fund for partial support of this research (award \#62195-DNI9) and to the US National Science Foundation (award \#2028617, CBET-FD).


\bibliography{references,references_houssem,final_references}

\end{document}